\documentclass[aps,twocolumn,superscriptaddress,showpacs,floatfix]{revtex4-1}
\usepackage[table]{xcolor}
\usepackage{amsfonts,amssymb,latexsym,xspace,epsfig,graphicx,color}
\usepackage[normalem]{ulem}
\usepackage{amsmath,enumerate,stmaryrd,xy,stackrel,multirow}
\usepackage[colorlinks=true,citecolor=blue,linkcolor=blue,urlcolor=blue]{hyperref}
\usepackage{subfigure}
\usepackage{tikz}
\usepackage{physics}
\usepackage{textpos}
\usepackage{qcircuit}
\usepackage{pifont}

\renewcommand{\H}{\mathcal{H}}
\newcommand{\bsc}{Barcelona Supercomputing Center (BSC), Barcelona 08034, Spain}
\newcommand{\icfo}{Institute of Photonic Sciences (ICFO), Castelldefels 08860, Spain}
\newcommand{\ub}{Departament de Física Quàntica i Astrofísica and Institut de Ciències del Cosmos, Universitat de Barcelona, Barcelona 08028, Spain}
\newcommand{\uab}{Departament de f\'isica, Universitat Aut\`onoma de Barcelona, Barcelona 08193, Spain}
\newcommand{\qilimanjaro}{Qilimanjaro Quantum Tech, Barcelona 08007, Spain}

\begin{document}
\title{Adiabatic quantum algorithm for artificial graphene}

\begin{abstract}
    
We devise a quantum-circuit algorithm to solve the ground state and ground energy of artificial graphene. 
The algorithm implements a Trotterized adiabatic evolution from a purely tight-binding Hamiltonian to one including kinetic, spin-orbit and Coulomb terms. 
The initial state is obtained efficiently using Gaussian-state preparation, while the readout of the ground energy is organized into
seventeen sets of measurements, irrespective of the size of the problem. 
The total depth of the corresponding quantum circuit scales polynomially with the size of the system. A full simulation of the algorithm is performed and ground energies are obtained for lattices with up to four hexagons. Our results are benchmarked with exact diagonalization for systems with one and two hexagons. For larger systems we use the exact statevector and approximate matrix product state simulation techniques. The latter allows to systematically trade off precision with memory and therefore to tackle larger systems. We analyze adiabatic and Trotterization errors, providing estimates for optimal periods and time discretizations given a finite accuracy. In the case of large systems we also study approximation errors.

\end{abstract}

\author{Axel P\'erez-Obiol}
\affiliation{\bsc}
\author{Adrián P\'erez-Salinas}
\affiliation{\bsc}
\author{Sergio S\'anchez-Ram\'irez}
\affiliation{\bsc}
\affiliation{\ub}
\author{Bruna G. M. Ara\'ujo}
\affiliation{\icfo}
\affiliation{\uab}
\author{Artur Garcia-Saez}
\affiliation{\bsc}
\affiliation{\qilimanjaro}

\maketitle

\section{Introduction}\label{sec:introduction}
Quantum computing is a celebrated emergent technology witnessing a surge of new developments both in theoretical and experimental sides during recent years~\cite{google_supremacy_2019, zhong_quantum_2020}. It has been used to address particular classical problems~\cite{grover, shor, brassard_amplitude_estimation_2002}, or more recently general optimization~\cite{Peruzzo_vqe_2014, farhi_qaoa_2014, optimization-lopez2015, optimization-rosenberg2016, optimization-moll2018} or Machine Learning~\cite{rebentrost_qsvm_2014, liu_rigorous_2020, schuld_circuit_2020, perezsalinas_data_2020, farhi_classification_2018, johri_nearest_2020, havlicek_supervised_2019}. However, quantum computing is particularly well suited for simulating quantum properties of nature~\cite{Cirac_simulation_2012, Childs_simulation_2018, Georgescu_simulation_2014}, while doing so in classical computers is extremely inefficient~\cite{Feynman_simulating_1982}. Some examples include quantum chemistry~\cite{mcardle_chemistry_2020, kandala_hardware_2017, hempel_quantum_2018, Nam_groundstate_2020, OMalley_scalable_2016}, condensed matter systems \cite{frey2022realization, doronin2021simulation}, or analysis of quantum data~\cite{huang_demonstrating_2021, huang_power_2021}. Despite this advantage, current capabilities of quantum hardware are still far from simulating systems of practical interest.

Adiabatic evolution processes are among the strategies that can be followed to perform computations on quantum devices. Adiabatic evolution, based on the adiabatic theorem~\cite{Kato1950adiabatic}, is implementable both in quantum-annealing and quantum gate-based~\cite{Albash_adiabatic_2018} hardware. In the adiabatic process, a time-dependent Hamiltonian interpolates from an initial well-controlled Hamiltonian to a final one encoding the problem of interest. Starting from the initial ground state, the system is gradually evolved according to the time-dependent Hamiltonian. For gapped systems and a slow enough transition, the final state approximates to the ground state of the problem Hamiltonian.

An interesting problem to solve with quantum computing is the simulation and analysis of materials down to the atomic level. In particular, single layer  crystalline solids allow to study interesting physical phenomena without the computational burden of full 3D material simulations. Among these, graphene stands out as a novel material with exciting mechanical, thermal, optical, and electronic properties~\cite{Geim_graphene_2007}, including  superconductivity~\cite{zhou_graphene_2021}.
Graphene is composed of a bidimensional lattice of carbon atoms displayed in a honeycomb structure. Its isolation and characterization in 2004 \cite{Novoselov2004graphene}, together with its demonstrated technological potential~\cite{GNassef2020grapheneerview}, has motivated the design and production of artificial graphene (AG)~\cite{Gibertini2009AG}, that is, more general two-dimensional fermionic platforms that maintain the hexagonal structure. 
AG systems preserve the main electronic properties of graphene,
owing to the preserved hexagonal symmetry of the lattice,
while representing
much more tunable platforms that permit to model imaginative electronic interactions or explore new phases of matter. A simple yet successful model to capture the electronic features of material lattices, including AG, is the Fermi-Hubbard model~\cite{fermi_hubbard}. This model considers hopping, Coulomb and spin-orbit interactions among the free-moving electrons in the lattice.

There already exist a number of approaches to solve this problem. 
It can be solved with high accuracy with exact diagonalization methods,
although the exponential scaling of the Hilbert space makes this approach
practically useless for already relatively small lattices.
Other classical methods such as DMRG~\cite{White_dmrg_1993, Schollwck_dmrg_2011} find an approximation to the ground state of a given system using variational approaches and composing smaller systems into a larger one. They can only be applied on classical machines since non-unitary operations are performed. Thus, the calculation is subject to a dimensionality curse in the description of the quantum state, and a trade-off between available accuracy and size arises. In the quantum side, variational methods are also explored~\cite{cade2020strategies}. Optimizing such problems is in general difficult and requires much classical computational power. 

In this work, we develop an algorithm for gate-based quantum computers that simulates an adiabatic evolution and outputs the ground state of the free-moving electrons in AG. The algorithm is composed of four steps. First, the problem is mapped from the fermionic operators to a qubit-based implementation suitable for quantum computers. The standard Jordan-Wigner map is used for this purpose. Second, an initial state to start the adiabatic process is generated. This can be done efficiently under some requirements~\cite{Jiang2018quantum}. Third, the adiabatic evolution is conducted. The recipe is based on that of a square lattice~\cite{cade2020strategies}, properly adapted and extended to match the present problem. Finally, a tailored measurement strategy is described to measure all terms in the AG Hamiltonian while minimizing the number of circuit executions. 

The algorithm here presented is efficient since its scaling in number of qubits and operations increases linearly with the size of the lattice, both for the preparation of the initial state and for each step of the adiabatic evolution. In the case of the measurement process, the number of measurements is constant as the system size increases. This scaling contrasts with the classical simulation of the same problem, where only the storing of the electronic wavefunction is exponential in the size of the system. In addition, note that no classical computational power is needed apart from processing data.

The quantum algorithm is tested using classical machines, including
the supercomputer MareNostrum 4, hosted at BSC~\cite{marenostrum}, in the case of large systems. The simulations are carried out using an exact statevector (SV) representation, provided by the {\tt Qibo} framework~\cite{Efthymiou2021qibo}, and a Matrix Product State (MPS) one. The latter is obtained by translating the quantum circuit to an MPS by the library {\tt quimb}~\cite{Gray2018quimb}, and then contracted using local truncation using the library {\tt RosNet} \cite{SanchezRamirez_rosnet_2021}.
Using exact simulation, the largest simulatable system is four hexagons in a $2\times 2$ lattice, with 32 qubits. In the MPS case, four hexagons in a $1 \times 4$ lattice, represented by 36 qubits, are achieved with high fidelities. The limits of these results are imposed by hardware.

The paper is structured as follows. In Sec.~\ref{sec:fermi_hubbard} the Fermi-Hubbard model and lattice definitions of AG are described. Sec.~\ref{sec:algorithm} is devoted to the recipe of all steps of the algorithm. Comments on the simulation part are detailed in Sec.~\ref{sec:simulation}. Results are given in Sec.~\ref{sec:results}. Further comments and conclusions can be read in Sec.~\ref{sec:conclusions}

\section{Fermi-Hubbard model graphene}\label{sec:fermi_hubbard}

The present work is based on the Fermi-Hubbard model~\cite{fermi_hubbard} as applied to a graphene lattice. The Fermi-Hubbard model is widely used in condensed matter as a reasonable approximation to fermionic systems~\cite{Jiang2018quantum}. It approximates the long-range Coulomb interactions among electrons in a crystal with local interactions in each lattice site. It includes a kinetic term modeling the hopping of electrons among first neighbours, and we also include the spin-orbit Rashba term, allowing electrons to hop while also switching the spin. The Hamiltonian of such model is defined as

\begin{align}\label{eq:h_fh}
\H_{FH} =& -t\sum_{\langle ij\rangle,\sigma}a_{i,\sigma}^\dagger a_{j,\sigma} +
U\sum_{i}a_{i,\uparrow}^\dagger a_{i,\uparrow} a_{i,\downarrow}^\dagger a_{i,\downarrow}
\nonumber\\&+\frac{2i}{3}\lambda_R
\sum_{\langle ij\rangle,\sigma,\sigma'}a_{i,\sigma'}^\dagger a_{j,\sigma}
[(\vec{\sigma}\times \vec{d}_{ij})_z]_{\sigma,\sigma'},
\end{align}
and for simplicity we define the hopping, Coulomb and spin-orbit Hamiltonians respectively as
\begin{eqnarray}
\H_H &=& -t\sum_{\langle ij\rangle,\sigma}a_{i,\sigma}^\dagger a_{j,\sigma}, \\
\H_C &=& U\sum_{i}a_{i,\uparrow}^\dagger a_{i,\uparrow} a_{i,\downarrow}^\dagger a_{i,\downarrow}, \\
\H_{SO} & = & \frac{2i}{3}\lambda_R
\sum_{\langle ij\rangle,\sigma,\sigma'}a_{i,\sigma'}^\dagger a_{j,\sigma}
[(\vec{\sigma}\times \vec{d}_{ij})_z]_{\sigma,\sigma'}.
\end{eqnarray}
$t, U, \lambda_R$ are respectively the hopping, Coulomb and spin-orbit couplings.
Operators $a_{i, \sigma}$ and  $a^\dagger_{i, \sigma}$ 
annihilate and create an electron
in site $i$ with spin $\sigma$.
Sums over $\langle i j \rangle$ in the case of hopping and spin-orbit terms take into consideration only the contributions between sites $(i, j)$ connected by a single edge in the graphene lattice, which we consider to have regular, non-periodic boundary conditions.
$\vec \sigma$ and $\vec d_{ij}$ in the spin-orbit term are the vector of Pauli matrices and the (unitary) vector between sites $i$ and $j$ in the lattice. Any length of $\vec d_{ij}$ can be absorbed in a simple re-scaling of $\lambda_R$ 

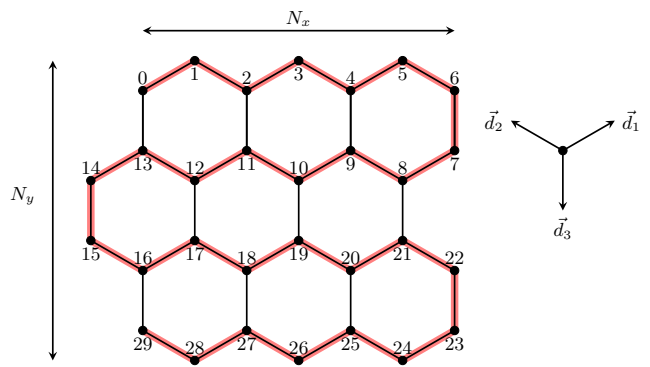
\begin{figure}[b]
    \centering
    \resizebox{\linewidth}{!}{
    \begin{tikzpicture}
    \foreach \x in {0,1,2}{
    \draw[red, line width=1.25mm, opacity=0.5] (\x*1.732,0) -- (\x*1.732 + 0.866,1/2);
    \draw[black, thick] (\x*1.732,0) -- (\x*1.732 + 0.866,1/2);
    \draw[red, line width=1.25mm, opacity=0.5] (\x*1.732 + 0.866,1/2) -- (\x*1.732 + 1.732,0);
    \draw[black, thick] (\x*1.732 + 0.866,1/2) -- (\x*1.732 + 1.732,0);
    \draw[red, line width=1.25mm, opacity=0.5] (\x*1.732,-1) -- (\x*1.732 + 0.866,-3/2);
    \draw[black, thick] (\x*1.732,-1) -- (\x*1.732 + 0.866,-3/2);
    \draw[red, line width=1.25mm, opacity=0.5] (\x*1.732 + 0.866,-3/2) -- (\x*1.732 + 1.732,-1);
    \draw[black, thick] (\x*1.732 + 0.866,-3/2) -- (\x*1.732 + 1.732,-1);}
    \draw[red, line width=1.25mm, opacity=0.5] (3*1.732,0) -- (3*1.732,-1);
    \foreach \x in {0,1,2, 3}{
    \draw[black, thick] (\x*1.732,0) -- (\x*1.732,-1);
    };
    \draw[red, line width=1.25mm, opacity=0.5] (-0.866,-3/2) -- (0,-1);
    \draw[black, thick] (-0.866,-3/2) -- (0,-1);
    \foreach \x in {0,1,2}{
    \draw[red, line width=1.25mm, opacity=0.5] (\x*1.732 - 0.866,-5/2) -- (\x*1.732,-3);
    \draw[red, line width=1.25mm, opacity=0.5] (\x*1.732,-3) -- (\x*1.732 + 0.866,-5/2);
    \draw[black, thick] (\x*1.732 - 0.866,-5/2) -- (\x*1.732,-3);
    \draw[black, thick] (\x*1.732,-3) -- (\x*1.732 + 0.866,-5/2);}
    \draw[red, line width=1.25mm, opacity=0.5] (0-.866,-3/2) -- (0 - 0.866,-5/2);
    \foreach \x in {0,1,2, 3}{
    \draw[black, thick] (\x*1.732-.866,-3/2) -- (\x*1.732 - 0.866,-5/2);
    }
    \foreach \x in {0,1,2, 3}{
    \draw[black, thick] (\x*1.732,0) -- (\x*1.732,-1);
    };
    \draw[red, line width=1.25mm, opacity=0.5] (3*1.732,-4) -- (3*1.732,-3);
    \foreach \x in {0,1,2, 3}{
    \draw[black, thick] (\x*1.732,-4) -- (\x*1.732,-3);
    };
    \draw[red, line width=1.25mm, opacity=0.5] (5 * .866, -5/2) -- (6 * .866, -3);
    \draw[black, thick] (5 * .866, -5/2) -- (6 * .866, -3);
    \foreach \x in {0,1,2}{
    \draw[red, line width=1.25mm, opacity=0.5] (\x*1.732,-4) -- (\x*1.732 + 0.866,-9/2);
    \draw[black, thick] (\x*1.732,-4) -- (\x*1.732 + 0.866,-9/2);
    \draw[red, line width=1.25mm, opacity=0.5] (\x*1.732 + 0.866,-9/2) -- (\x*1.732 + 1.732,-4);
    \draw[black, thick] (\x*1.732 + 0.866,-9/2) -- (\x*1.732 + 1.732,-4);}
    \foreach \x in {0,2,4, 6}{
    \filldraw[black] (\x*.866,0) circle (2pt)
    node[anchor=south]{\x};
    };
    \foreach \x in {1,3,5}{
    \filldraw[black] (\x*.866,1/2) circle (2pt) node[anchor=north]{\x};
    };
    \foreach \l [count=\i, evaluate=\i as \x using int(2*\i - 2)]in {13,11,9, 7}{
    \filldraw[black] (\x*.866,-1) circle (2pt) node[anchor=north]{\l};
    };
    \foreach \l [count=\i, evaluate=\i as \x using int(2*\i - 3)] in {14, 12,10,8}{
    \filldraw[black] (\x*.866,-3/2) circle (2pt) node[anchor=south]{\l};
    };
    \foreach \l  [count=\i, evaluate=\i as \x using int(2*\i - 3)] in {15, 17,19,21}{
    \filldraw[black] (\x*.866,-5/2) circle (2pt) node[anchor=north]{\l};
    };
    \foreach \l [count=\i, evaluate=\i as \x using int(2*\i - 2)] in {16,18,20, 22}{
    \filldraw[black] (\x*.866,-3) circle (2pt) node[anchor=south]{\l};
    };
    \foreach \l [count=\i, evaluate=\i as \x using int(2*\i - 2)] in {29,27,25, 23}{
    \filldraw[black] (\x*.866,-4) circle (2pt) node[anchor=north]{\l};
    };
    \foreach \l [count=\i, evaluate=\i as \x using int(2*\i - 1)]in {28, 26,24}{
    \filldraw[black] (\x * .866,-9/2) circle (2pt) node[anchor=south]{\l};
    };
    \draw[black, thick, stealth-stealth] (0,1) -- (3*1.732,1);
    \node at (1.5*1.732,1.2) {$N_x$};
    \draw[black, thick, stealth-stealth] (-1.5,1/2) -- (-1.5,-9/2);
    \node at (-2,-7/4) {$N_y$};
    \filldraw[black] (7, -1) circle (2pt);
    \draw[black, thick, -stealth] (7, -1) -- (7 + .866,-1/2) node[anchor=west]{$\vec d_1$};
    \draw[black, thick, -stealth] (7, -1) -- (7 - .866,-1/2) node[anchor=east]{$\vec d_2$};
    \draw[black, thick, -stealth] (7, -1) -- (7, -2) node[anchor=north]{$\vec d_3$};
\end{tikzpicture}}
    \caption{Graphene lattice used in this work, defined by the number of hexagons in each dimension $(N_x,N_y)$. Sites are numbered following the red shadow. The regular hexagons are composed of the unitary vectors connecting sites $\vec d_i$, on the right side, defined in Eq.~\eqref{eq:unitary_vectors}.
    }
    \label{fig:lattice}
\end{figure}

The graphene lattice here considered is depicted in Fig.~\ref{fig:lattice}, with each site corresponding to two orbitals, one for spin up $\uparrow$, and another for spin down, $\downarrow$. The lattice follows a standard honeycomb structure where the number of hexagons on directions $x$ and $y$ is always maintained. All hexagons are regular, with all sites equally spaced, and oriented in a zig-zag configuration. The vectors connecting sites are easily found as
\begin{equation}\label{eq:unitary_vectors}
\begin{matrix}
\vec{d}_1 &=& \frac{\sqrt{3}}{2} \hat{e}_x + \frac{1}{2} \hat{e}_y \\
\vec{d}_2 &=& -\frac{\sqrt{3}}{2} \hat{e}_x + \frac{1}{2} \hat{e}_y \\
\vec{d}_1 &=& -\hat{e}_y,
\end{matrix}
\end{equation}
where $\hat{e_j}$ is the unit vector in direction $j$.

\section{Description of the algorithm}\label{sec:algorithm}

In this work we explore the implementation of the Fermi-Hubbard model on a quantum computer, that is, a native simulation framework. The algorithm goes as follows. First, one needs to map the problem to a quantum computer, in this case using the Jordan-Wigner mapping.
Second, an easy-to-prepare state is generated, as explained in~\cite{Jiang2018quantum}. The initial state is later subject to adiabatic evolution until a good approximation of the target state is achieved. After the  evolution is performed, a change of basis is applied to measure the final energy in the most possible efficient way.

\subsection{Jordan-Wigner mapping}

The first step in the algorithm is to map the sites and orbitals in the Fermi-Hubbard model to qubits in a quantum computer.
The correspondence between orbitals and qubits traces the snake-like pattern shadowed in red in Fig.~\ref{fig:lattice}, with two qubit numbers for each site, since each orbital can be occupied by an electron with spin $\uparrow$ or $\downarrow$.
The particular relation between spins and qubit numbers depends on the row of the lattice. We follow the numbering from Fig.~\ref{fig:lattice}, and label rows and columns of qubits from top to bottom and left to right, starting with zero. Thus, in even rows of sites, the spins $\uparrow$ have an even label, while $\downarrow$ have an odd one. In odd rows, the label is inverted.

\begin{eqnarray}
s_{\rm even\, rows} & \rightarrow & 2s, 2s + 1 = \{\uparrow, \downarrow\} \\
s_{\rm odd\, rows} & \rightarrow & 2s, 2s + 1 = \{\downarrow, \uparrow\} 
\end{eqnarray}
For example, in the lattice of Fig.~\ref{fig:lattice}, spins $\uparrow$ and $\downarrow$
of site number six, correspond to labels 12 and 13, while the same spins for site
number seven are mapped to qubit numbers 15 and 14.

Each combination of site and spin is then mapped to qubits via the Jordan-Wigner mapping~\cite{Nielsen2005JW}. According to this mapping, the creation and annihilation operators are transformed via
\begin{eqnarray}
a^\dagger_j = \prod_{k=0}^{j - 1} \sigma^{(z)}_k\frac{1}{2} (\sigma^{(x)}_j + i \sigma^{(y)}_j) \\
a_j =  \prod_{k=0}^{j - 1} \sigma^{(z)}_k \frac{1}{2} (\sigma^{(x)}_j - i \sigma^{(y)}_j) \\
\end{eqnarray}
where the chain of operators $ \prod_{k=0}^{j - 1} \sigma^{(z)}_k$ is used to keep track of the relative signs under commutations of two fermions.

\subsection{Initial state preparation}

The initial state used for the adiabatic evolution will be the ground state of a tight binding Hamiltonian $\H_{TB}$ defined as 
\begin{equation}\label{eq:h_tb}
    \H_{TB} = \H_H + \H_{SO}, 
\end{equation}
that is, a Hamiltonian where the only energetic terms are among the sites. Alternatively, it is also possible to take the hopping Hamiltonian $\H_H$ as the starting point.

Both kinds of Hamiltonian satisfy the conditions to be a fermionic Gaussian state~\cite{Jiang2018quantum, Bach1994generalized} defined as
\begin{equation}
    \H_G = \sum_{j, k=1}^N M_{jk} a^\dagger_j a_k + \sum_{j, k=1}^N (\Delta_{jk} a^\dagger_j a^\dagger_k  + h.c.).
\end{equation}

Jiang et al. in Ref.~\cite{Jiang2018quantum} show that the ground state of an arbitrary Gaussian Hamiltonian can be constructed efficiently on a quantum computer.
In particular, it requires $\mathcal{O}(n^2)$ operations and $\mathcal{O}(n)$ depth, where $n$ is the total number of qubits. The scaling for the problem here considered will be $n =\mathcal{O}(N_x N_y)$. In this work, the {\tt Openfermion}~\cite{McClean2020openfermion} implementation for que quantum computing library {\tt Cirq}~\cite{developers2021cirq} is used to construct this ground state.

Notice that the mapping used in this algorithm follows a 1D path, while the problem is 2D. This is a consequence of the fermionic features of the model. All interchanges of two fermions must carry a phase change. To the best of our knowledge, there is no 2D fermionic mapping suiting this problem, apart from the folding of a 1D chain into a larger dimensional space.

\subsection{Adiabatic Evolution}

Adiabatic evolution is the central piece of the algorithm here developed. In the adiabatic evolution the system is initially affected by a Hamiltonian whose ground state is known. Then, the Hamiltonian is slowly changed to another one encoding the problem to be solved and whose ground state is unknown. If the evolution is performed slowly enough, then the final state will be close to the ground state, the solution of the problem, of the final Hamiltonian.~\cite{Born1928adiabatic, Kato1950adiabatic, Avron1999adiabatic}.
The evolution time must be at least $\mathcal{O}(g^2)$ where $g$ is the gap between the ground and first-excited states~\cite{Born1928adiabatic}.

In our case, the initial Hamiltonian and ground state are $\H_{TB}$ and $\ket{\psi_{TB}}$. The problem Hamiltonian is $\H_{FH}$, with ground state $\ket{\psi_{FH}}$. The time-depending Hamiltonian interpolating between both extrema during a total evolution time $T$ is
\begin{equation}\label{eq:time_ham}
    \H(s) = (1 - s) \H_{TB} + s \H_{FH}, \qquad s = t/T.
\end{equation}
By direct substitution from Eqs.~\eqref{eq:h_fh} and~\eqref{eq:h_tb} it is possible to rewrite the adiabatic time depending Hamiltonian as
\begin{equation}
    \H(s) = \H_H + \H_{SO} + s\H_C,
\end{equation}
that is, the Coulomb term is adiabatically turned on.

To perform the state evolution, the Schr\"odinger equation with the above time-dependent Hamiltonian is applied,
\begin{equation}
    \ket{\psi(T)} = e^{-i\int_0^T \H(t) dt} \ket{\psi(0)}.
\end{equation}
This formal solution can be approximated by discretizing the evolution path as
\begin{equation}\label{eq:adiabatic}
    \ket{\psi(T)} \approx \prod_{j=0}^N e^{-i\H(j\delta t) \delta t} \ket{\psi(0)},
\end{equation}
where $T$ is the total evolution time and $\delta t = T/N$ is the time taken in each step. The final state $\ket{\psi(T)}$ is approximated with an error in the relative fidelity of $\delta t^2$~\cite{Lloyd1996universal}. In this work, this evolution is referred to as {\sl adiabatic evolution}, and its error as {\sl adiabatic error}.

\subsubsection*{Adiabatic evolution on a quantum computer}

The goal of the algorithm is to run on a quantum computer. Such computers can perform only a set of quantum operations, or quantum gates, on the states. The set of gates is chosen to conform a universal set of gates capable to perform any operation. A possible attempt to execute the adiabatic evolution on a quantum computer is to compile the gate $e^{-i\H(\delta t j)\delta t}$ to be decomposed into available operations~\cite{botea2018complexity}. This process is however not explored in this work.

The recipe followed in this work is the Trotterization of each adiabatic step~\cite{childs2021trotter} to obtain an approximate evolution that can be directly applied on the quantum circuit. In this case, Trotterization, already explored in Eq.~\eqref{eq:adiabatic}, consists in applying each term in the Hamiltonian separately.
Its general form is 
\begin{equation}
    \begin{matrix}
    \H & = & \H_1 + \H_2 \\
     & \Downarrow & \\
    e^{-i\H t} & = & e^{-i\H_1 t}e^{-i\H_2 t} + \mathcal{O}([\H_1, \H_2], t^2)
    \end{matrix}.
\end{equation}
The present problem uses Trotterization to decompose the evolution into pieces that fit in the quantum computer. Those pieces are
\begin{widetext}\begin{equation}\label{eq:trotter}
    e^{-i\H(s)\delta t} \approx
    \underbrace{\prod_{i, \sigma}e^{-iU s \delta_t a^\dagger_{i, \uparrow} a^\dagger_{i, \downarrow}a_{i, \uparrow} a_{i, \downarrow}}}_{\exp(-i\H_C s \delta t)} \times
    \underbrace{\prod_{\langle i j\rangle, \sigma \sigma'}e^{it \delta_t a^\dagger_{i, \sigma} a_{j, \sigma}}}_{\exp(-i\H_H \delta t)}
    \times \underbrace{\prod_{\langle i j\rangle, \sigma}e^{-i\frac{2i \lambda_R}{3} \delta_t a^\dagger_{i, \sigma} a_{j, \sigma'}[(\vec{\sigma}\times \vec{d}_{ij})_z]_{\sigma,\sigma'}}}_{\exp(-i\H_{SO} \delta t)}.
\end{equation}\end{widetext}

Notice that in the previous adiabatic case the Trotterization is done by splitting the full evolution into time steps. In the circuit evolution, each of these time steps is further divided into one step per term in the Hamiltonian. This way, the algorithm is decomposed into small pieces that can be directly applied on quantum computers, involving only one- and two-qubit operations, without any further compilation, up to trivial changes of gates to the native operations of the processor of interest. 

The Coulomb term contributes with as many operations as sites, $(2N_x + 2)(N_y + 1) - 2$, and hopping and spin-orbit terms count edges in the lattice, $(2N_x + 1)(N_y + 1) - 2 + (N_x + 1)N_y = 3N_xN_y + 2(N_x + N_y) + 1$. Thus, the number of operations in the circuit Trotterization is $\mathcal{O}(N_xN_y)$ larger than the standard adiabatic evolution. As a trade-off, each operation is less costly. The error in the circuit Trotterization is however not strongly affected by the size of the system. Since all interactions are 2-local, a first order Trotter path is affected only by commutators of first neighbors~\cite{childs2021trotter}. However, as more sites are added to the system, more of them have 3 neighbor sites, increasing the overall error.

This circuit Trotterization requires to implement all the interactions present in the FH Hamiltonian.
However, the Jordan-Wigner mapping, which translates the 2D fermionic lattice
into a 1D qubit string, implies Hamiltonian terms operate in non-neighbouring qubits.
To circumvent this limitation, a simple and inefficient strategy would be to add swap gates among the corresponding qubits to set a given pair together and perform the adiabatic evolution term on adjacent qubits. Then the qubits are restored to its original position. A more efficient implementation of this idea involves more sophisticated swapping schemes~\cite{Kivlichan2018simulation, cade2020strategies} to minimize the number of operations required for performing a full evolution step for a particular lattice. In the present work, this strategy is modified to address the hexagonal lattice problem. 

\begin{figure}[b!]
\centering
    \resizebox{.8\linewidth}{!}{
    \begin{tikzpicture}
    \foreach \x in {0,1,2,3,4,5,6,7}{
    \node[circle,draw] (c) at (\x,0){\x}; 
    };
    \foreach \x\l in {0/1,1/0,2/3,3/2,4/5,5/4,6/7,7/6}{
    \node[circle,draw] (c) at (\x,-1){\l}; 
    };
    \foreach \x\l in {0/1,1/3,2/0,3/5,4/2,5/7,6/4,7/6}{
    \node[circle,draw] (c) at (\x,-2){\l}; 
    };
    \foreach \x\l in {0,2,4,6}{
    \draw[black, thick, -stealth] (\x + 0.25, -0.25) -- (\x + 0.75, -0.75);
    \draw[black, thick, -stealth] (\x + 0.75, -0.25) -- (\x + 0.25, -0.75);
    };
    \foreach \x\l in {1,3,5}{
    \draw[black, thick, -stealth] (\x + 0.25, -1.25) -- (\x + 0.75, -1.75);
    \draw[black, thick, -stealth] (\x + 0.75, -1.25) -- (\x + 0.25, -1.75);
    };
    \draw[black, thick, -stealth] (0, -1.35) -- (0, -1.70);
    \draw[black, thick, -stealth] (7, -1.35) -- (7, -1.70);
\end{tikzpicture}}
\caption{Swap operations performed on the qubits belonging to the same row. First, the pairs $(2k, 2k + 1), k \in \mathbb{N}$ are swapped. Then, the pairs $(2k + 1, 2k + 2), k \in \mathbb{N}$. This permutation is cyclic and after as many steps as qubits in the chain the original state is recovered. Along the path, all operations in $\H_{FH}$ can be applied between adjacent qubits at least once. }
\label{fig:swap_step}
\end{figure}
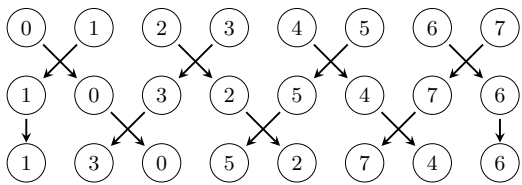

To perform all operations, all required combinations of qubits must be set together at some point in the algorithm. For this, the full qubit chain is divided into $N_y + 1$ sub-chains, each representing one row of sites from the lattice. Each row has $2N_x + 2$ sites, except for the first and the last ones, having just $2N_x + 1$. Accordingly, each qubit chain has twice the number of sites.
The evolution steps can only be applied between adjacent qubits. For Hamiltonian vertical hopping terms, connecting qubits from different rows, this can only happen by swapping qubit states and effectively bringing them to the last and first qubits of adjacent rows. In these positions, corresponding, for example, to sites 6 and 7 or 14 and 15 in Fig.~\ref{fig:lattice}, is where the actual evolution operators described above are applied at each evolution step. Series of fermionic swap gates are added to make all qubits iteratively  go through the positions of interest. Fermionic swaps, instead of regular swaps, are required to maintain the correct parity. Their operator and matrix representation reads
\begin{equation}
\small
{\rm FSWAP} = 
1+a^\dagger_0 a_1 + a^\dagger_1 a_0
-a^\dagger_0 a_0 - a^\dagger_1 a_1=
\\ \begin{pmatrix}
    1 & 0 &  0 &0 \\ 0 & 0 & 1 &0 \\ 0 & 1 &  0 &0 \\ 0 & 0 &  0 & -1 \\
    \end{pmatrix}.
\end{equation}

The fermionic swaps are composed of the iteration described in Fig.~\ref{fig:swap_step}. First, fermionic swaps between even-odd qubits pairs $(2k, 2k + 1),\; k\in \mathbb{N}$ are applied. Then, a fermionic swap among odd-even pairs $(2k + 1, 2k + 2),\; k\in \mathbb{N}$ follows. The permutation, applied separately in each row,  satisfies the following properties. First, it is periodical when applied repeatedly on the same qubit string with periodicity equal to the length of the string. Second, all possible even-even and odd-odd combinations appear in the first and last pair of qubits in the chain. Finally, the first and last qubits receive all possible values iteratively. These properties allow to execute the full Hamiltonian only on a very restricted portion of the chain. Notice that due to the periodicity of the aforementioned permutation, the scaling of any algorithm based on this scheme is at most $\mathcal{O}(N_x)$.

\begin{table}[t!]
    \flushleft
    \begin{tabular}{|c c c c c c c c|}\hline
        x & x & 0 & 1 & 2 & 3 & 4 & 5 \\
        13 & 12 & 11 & 10 & 9 & 8 & 7 & 6 \\
        14 & 15 & 16 & 17 & 18 & 19 & x & x \\ \hline
        x & x & 1 & 3 & 0 & 5 & 2 & \cellcolor{green!50!} 4 \\
        \cellcolor{green!50!}12 & 10 & 13 & 8 & 11 & 6 & 9 & \cellcolor{green!50!}7 \\
        \cellcolor{green!50!}15 & 17 & 14 & 19 & 16 & 18 & x & x \\ \hline
        x & x & 3 & 5 & 1 & 4 & 0 & 2 \\
        10 & 8 & 12 & 6 & 13 & 7 & 11 &  9 \\
        17 & 19 & 15 & 18 & 14 & 16 & x & x \\ \hline
        x & x & 5 & 4 & 3 & 2 & 1 & \cellcolor{green!50!}0 \\
        \cellcolor{green!50!}8 & 6 & 10 & 7 & 12 & 9 & 13 & \cellcolor{green!50!}11 \\
        \cellcolor{green!50!}19 & 18 & 17 & 16 & 15 & 14 & x & x \\ \hline
        x & x & 5 & 4 & 3 & 2 & 1 & 0 \\
        6 & 7 & 8 & 9 & 10 & 11 & 12 & 13 \\
        19 & 18 & 17 & 16 & 15 & 14 & x & x \\ \hline
        x & x & 5 & 4 & 3 & 2 & 1 & 0 \\
        7 & 9 & 6 & 11 & 8 & 13 & 10 & 12 \\
        19 & 18 & 17 & 16 & 15 & 14 & x & x \\ \hline
        x & x & 4 & 2 & 5 & 0 & 3 & \cellcolor{green!50!}1 \\
        \cellcolor{green!50!}9 & 11 & 7 & 13 & 6 & 12 & 8 & \cellcolor{green!50!}10 \\
        \cellcolor{green!50!}18 & 16 & 19 & 14 & 17 & 15 & x & x \\ \hline
        x & x & 2 & 0 & 4 & 1 & 5 & 3 \\
        11 & 13 & 9 & 12 & 7 & 10 & 6 &  8 \\
        16 & 14 & 18 & 15 & 19 & 17 & x & x \\ \hline
        x & x & 0 & 1 & 2 & 3 & 4 & \cellcolor{green!50!}5 \\
        \cellcolor{green!50!}13 & 12 & 11 & 10 & 9 & 8 & 7 & \cellcolor{green!50!}6 \\
        \cellcolor{green!50!}14 & 15 & 16 & 17 & 18 & 19 & x & x \\ \hline
    \end{tabular} 
    \begin{textblock}{3.5}(3,-5.1)
    $\rightarrow$~Horizontal interaction $(2,4)$
    \end{textblock}
    \begin{textblock}{3.5}(3,-3.7)
    $\rightarrow$~Vertical interaction $(0,11)$
    \end{textblock}
    \begin{textblock}{3.5}(3,-3)
    $\rightarrow$~Extra permutation step
    \end{textblock}
    \begin{textblock}{8}(-0.025,-5.12)
    \hspace{1mm}\begin{tikzpicture}
    \draw[red] (2.8, 7.45) rectangle (3.6, 7.75);
    \draw[red] (0, 7.1) rectangle (0.8, 7.4);
    \draw[red] (0, 6.7) rectangle (0.8, 7);
    
    \draw[red] (2.8, 6.35) rectangle (3.6, 6.65);
    \draw[red] (0, 5.95) rectangle (0.8, 6.25);
    \draw[red] (0, 5.6) rectangle (0.8, 5.9);
    
    \draw[red] (0, 4.9) rectangle (0.8, 5.2);
    
    \draw[red] (0, 2.6) rectangle (0.8, 2.9);
    
    \draw[red] (2.8, 1.85) rectangle (3.6, 2.15);
    \draw[red] (0, 1.5) rectangle (0.8, 1.8);
    \draw[red] (0, 1.15) rectangle (0.8, 1.45);
    
    \draw[red] (2.8, 0.75) rectangle (3.6, 1.05);
    \draw[red] (0, 0.4) rectangle (0.8, 0.7);
    \draw[red] (0, 0) rectangle (0.8, 0.3);
    \end{tikzpicture}
    \end{textblock}
    \caption{Steps taken to implement one step of the trotterized evolution for the hopping Hamiltonian $\H_H$. Each number corresponds to its qubit in the chain. Qubits inside a red frame suffer an horizontal interaction while qubits in green are affected by a vertical one. In the middle of the algorithm, the middle row must apply extra permutation steps to catch up with the other rows.}
    \label{tab:step_trotter}
\end{table}

Each operator in the Hamiltonian $\H_{FH}$, is translated easily into a two-qubit operation, where both qubits are adjacent. In the Coulomb case, the interactions involved only measure the occupation level of the orbitals, adding it to the energy if both electrons $(\uparrow, \downarrow)$ are present in the same orbital, thus
\begin{equation} 
h_C = a^\dagger_0a^\dagger_1 a_0 a_1 = \begin{pmatrix}
    0 & 0 &  0 &0 \\ 0 & 0 &  0 &0 \\ 0 & 0 &  0 &0 \\ 0 & 0 &  0 & 1 \\
    \end{pmatrix}\end{equation}  
with evolution operator
\begin{equation}\label{eq:op_coulomb}
U_C(\theta) = e^{-iUh_C\,\theta} = \begin{pmatrix}
    1 & 0 &  0 & 0 \\ 0 & 1 &  0 &0 \\ 0 & 0 &  1 &0 \\ 0 & 0 &  0 & e^{-iU \theta} \\
    \end{pmatrix}.
    \end{equation} 
For the hopping case, the Hamiltonian triggers a movement from one site to another without changing the spin. This interaction appears between qubits whose labels are even-even/odd-odd in the same row or even-odd/odd-even in different rows.
The corresponding matrix is
    \begin{equation}
h_H = a^\dagger_0 a_1 + a^\dagger_1 a_0 = \begin{pmatrix}
    0 & 0 &  0 &0 \\ 0 & 0 & 1 &0 \\ 0 & 1 &  0 &0 \\ 0 & 0 &  0 & 0
    \end{pmatrix},
\end{equation}
while its evolution is given by the operation
\begin{equation}\label{eq:op_hopping}
U_{H}(\theta) = e^{-ith_H \, \theta} = \begin{pmatrix}
    1 & 0 &  0 &0 \\ 0 & \cos(t \theta) & -i \sin(t \theta) &0 \\ 0 & -i \sin(t \theta) & \cos(t \theta) &0 \\ 0 & 0 &  0 & 1 \\
    \end{pmatrix}.
\end{equation}

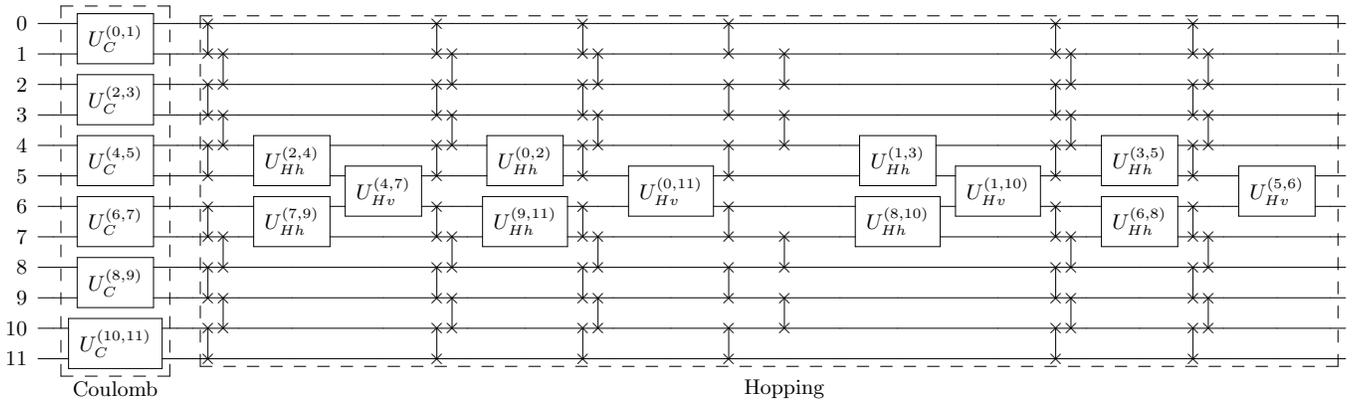
\begin{figure*}
\resizebox{\linewidth}{!}{
\hspace{0.5cm}\Qcircuit @C=0.7em @R=0.5em{
\lstick{0 } & \qw & \multigate{1}{U_C^{(0, 1)}}   & \qw & \qw & \qswap      & \qw     & \qw &  \qw                           &  \qw                           & \qswap      & \qw     & \qw  &  \qw                            & \qswap      & \qw     & \qw  &  \qw                            & \qswap      & \qw     & \qw  &  \qw                            &  \qw                            & \qswap      & \qw     & \qw  &  \qw                            & \qswap      & \qw     & \qw &  \qw                            & \qw & \qw \\
\lstick{1 } & \qw & \ghost{U_C^{(0, 1)}}          & \qw & \qw & \qswap \qwx & \qswap      & \qw &  \qw                           &  \qw                           & \qswap \qwx & \qswap      & \qw  &  \qw                            & \qswap \qwx & \qswap      & \qw  &  \qw                            & \qswap \qwx & \qswap      & \qw  &  \qw                            &  \qw                            & \qswap \qwx & \qswap      & \qw  &  \qw                            & \qswap \qwx & \qswap      & \qw &  \qw                            & \qw & \qw \\
\lstick{2 } & \qw & \multigate{1}{U_C^{(2, 3)}}   & \qw & \qw & \qswap      & \qswap \qwx & \qw &  \qw                           &  \qw                           & \qswap      & \qswap \qwx & \qw  &  \qw                            & \qswap      & \qswap \qwx & \qw  &  \qw                            & \qswap      & \qswap \qwx & \qw  &  \qw                            &  \qw                            & \qswap      & \qswap \qwx & \qw  &  \qw                            & \qswap      & \qswap \qwx & \qw &  \qw                            & \qw & \qw \\
\lstick{3 } & \qw & \ghost{U_C^{(2, 3)}}         & \qw & \qw & \qswap \qwx & \qswap      & \qw &  \qw                           &  \qw                           & \qswap \qwx & \qswap      & \qw  &  \qw                            & \qswap \qwx & \qswap      & \qw  &  \qw                            & \qswap \qwx & \qswap      & \qw  &  \qw                            &  \qw                            & \qswap \qwx & \qswap      & \qw  &  \qw                            & \qswap \qwx & \qswap      & \qw &  \qw                            & \qw & \qw \\
\lstick{4 } & \qw & \multigate{1}{U_C^{(4, 5)}}   & \qw & \qw & \qswap      & \qswap \qwx & \qw & \multigate{1}{U_{Hh}^{(2, 4)}} &  \qw                           & \qswap      & \qswap \qwx & \qw  & \multigate{1}{U_{Hh}^{(0, 2)}}  & \qswap      & \qswap \qwx & \qw  &  \qw                            & \qswap      & \qswap \qwx & \qw  & \multigate{1}{U_{Hh}^{(1, 3)}}  &  \qw                            & \qswap      & \qswap \qwx & \qw  & \multigate{1}{U_{Hh}^{(3, 5)}}  & \qswap      & \qswap \qwx & \qw &  \qw                            & \qw & \qw \\
\lstick{5 } & \qw & \ghost{U_C^{(4, 5)}}          & \qw & \qw & \qswap \qwx & \qw         & \qw & \ghost{U_{Hh}^{(2, 4)}}        & \multigate{1}{U_{Hv}^{(4, 7)}} & \qswap \qwx & \qw         & \qw  & \ghost{U_{Hh}^{(0, 2)}}         & \qswap \qwx & \qw         & \qw  & \multigate{1}{U_{Hv}^{(0, 11)}} & \qswap \qwx & \qw         & \qw  & \ghost{U_{Hh}^{(1, 3)}}         & \multigate{1}{U_{Hv}^{(1, 10)}} & \qswap \qwx & \qw         & \qw  & \ghost{U_{Hh}^{(3, 5)}}         & \qswap \qwx & \qw         & \qw & \multigate{1}{U_{Hv}^{(5, 6)}}  & \qw & \qw \\
\lstick{6 } & \qw & \multigate{1}{U_C^{(6, 7)}}   & \qw & \qw & \qswap      & \qw         & \qw & \multigate{1}{U_{Hh}^{(7, 9)}} & \ghost{U_{Hv}^{(4, 7)}}        & \qswap      & \qw         & \qw  & \multigate{1}{U_{Hh}^{(9, 11)}} & \qswap      & \qw         & \qw  & \ghost{U_{Hv}^{(0, 11)}}        & \qswap      & \qw         & \qw  & \multigate{1}{U_{Hh}^{(8, 10)}} & \ghost{U_{Hv}^{(1, 10)}}        & \qswap      & \qw         & \qw  & \multigate{1}{U_{Hh}^{(6, 8)}}  & \qswap      & \qw         & \qw & \ghost{U_{Hv}^{(5, 6)}}         & \qw & \qw \\
\lstick{7 } & \qw & \ghost{U_C^{(6, 7)}}          & \qw & \qw & \qswap \qwx & \qswap      & \qw & \ghost{U_{Hh}^{(7, 9)}}        &  \qw                           & \qswap \qwx & \qswap      & \qw  & \ghost{U_{Hh}^{(9, 11)}}        & \qswap \qwx & \qswap      & \qw  &  \qw                            & \qswap \qwx & \qswap      & \qw  & \ghost{U_{Hh}^{(8, 10)}}        &  \qw                            & \qswap \qwx & \qswap      & \qw  & \ghost{U_{Hh}^{(6, 8)}}         & \qswap \qwx & \qswap      & \qw &  \qw                            & \qw & \qw \\
\lstick{8 } & \qw & \multigate{1}{U_C^{(8, 9)}}   & \qw & \qw & \qswap      & \qswap \qwx & \qw &  \qw                           &  \qw                           & \qswap      & \qswap \qwx & \qw  &  \qw                            & \qswap      & \qswap \qwx & \qw  &  \qw                            & \qswap      & \qswap \qwx & \qw  &  \qw                            &  \qw                            & \qswap      & \qswap \qwx & \qw  &  \qw                            & \qswap      & \qswap \qwx & \qw &  \qw                            & \qw & \qw \\
\lstick{9 } & \qw & \ghost{U_C^{(8, 9)}}          & \qw & \qw & \qswap \qwx & \qswap      & \qw &  \qw                           &  \qw                           & \qswap \qwx & \qswap      & \qw  &  \qw                            & \qswap \qwx & \qswap      & \qw  &  \qw                            & \qswap \qwx & \qswap      & \qw  &  \qw                            &  \qw                            & \qswap \qwx & \qswap      & \qw  &  \qw                            & \qswap \qwx & \qswap      & \qw &  \qw                            & \qw & \qw \\
\lstick{10} & \qw & \multigate{1}{U_C^{(10, 11)}} & \qw & \qw & \qswap      & \qswap \qwx & \qw &  \qw                           &  \qw                           & \qswap      & \qswap \qwx & \qw  &  \qw                            & \qswap      & \qswap \qwx & \qw  &  \qw                            & \qswap      & \qswap \qwx & \qw  &  \qw                            &  \qw                            & \qswap      & \qswap \qwx & \qw  &  \qw                            & \qswap      & \qswap \qwx & \qw &  \qw                            & \qw & \qw \\
\lstick{11} & \qw & \ghost{U_C^{(10, 11)}}        & \qw & \qw & \qswap \qwx & \qw         & \qw &  \qw                           &  \qw                           & \qswap \qwx & \qw         & \qw  &  \qw                            & \qswap \qwx & \qw         & \qw  &  \qw                            & \qswap \qwx & \qw         & \qw  &  \qw                            &  \qw                            & \qswap \qwx & \qw         & \qw  &  \qw                            & \qswap \qwx & \qw         & \qw &  \qw                            & \qw & \qw
 \\
 & & \push{\rm Coulomb} & & & & & & & & & & & & & & & & & \push{\rm Hopping}
\protect\gategroup{1}{3}{12}{3}{.7em}{--}
\protect\gategroup{1}{6}{12}{32}{.7em}{--}
}}
\caption{Schematic description of an adiabatic step of the circuit, for the case of the Coulomb and hopping terms, for a single hexagon. In the Coulomb part, on the left, the gates $U_C^{(q_1, q_2)}$ perform the evolution between qubits $q_1, q_2$, corresponding to the same site in the lattice. The parameter $\theta$, not explicitly written, in this gate as defined in Eq.~\eqref{eq:op_coulomb} will depend in the exact step of the evolution. In the hopping term, on the right, the fermionic-swapping layers are interspersed with operations $U_{Hh}^{(q_1, q_2)}, U_{Hv}^{(q_1, q_2)}$ connecting qubits $q_1, q_2$ through a horizontal (h) or vertical (v) edge. The operations are labelled according to the representation qubits, and not to the circuit qubits, which are always the same ones. In the hopping case, the parameter $\theta$ of $U_H$ as defined in Eq.~\eqref{eq:op_hopping} is constant for all adiabatic steps.}
\label{fig:gates_adiabatic}
\end{figure*}

In the spin-orbit case, the Hamiltonian allows electrons to hop between lattice sites while flipping the spin. This interaction takes place between even-odd/odd-even qubit pairs in the same row and even-even/odd-odd pairs in different rows. It also carries a complex phase $\varphi$ whose exact value depends on the spins involved and the lattice vector $\vec d_i$. For adjacent qubits, the operation is
\begin{equation}
\small
h_{SO} = \frac{2i}{3}\lambda_R \left(e^{i\varphi}a^\dagger_0 a_1 - e^{-i\varphi}a^\dagger_1 a_0\right) = \\ \begin{pmatrix}
    0 & 0 &  0 &0 \\ 0 & 0 & ie^{i\varphi} &0 \\ 0 & -ie^{-i\varphi} &  0 &0 \\ 0 & 0 &  0 & 0 \\
    \end{pmatrix}
\end{equation}
and its corresponding evolution is
\begin{multline}\label{eq:op_rashba}
    U_{SO}(\theta) = e^{-i2/3\lambda_R h_{SO}\, \theta} = \\ \begin{pmatrix}
    1 & 0 &  0 &0 \\ 0 & \cos(2/3\lambda_R\theta) & -e^{i\varphi} \sin(2/3\lambda_R\theta) &0 \\ 0 & e^{-i\varphi} \sin(2/3\lambda_R\theta) & \cos(2/3\lambda_R\theta) &0 \\ 0 & 0 &  0 & 1 \\
    \end{pmatrix}.
\end{multline}

Each piece of the Hamiltonian $\H_{FH}$ will be applied separatedly. Coulomb terms only affect adjacent pairs of qubits, so there is no need to apply any further transformation. For the hopping terms, the full chain permutation must be applied. All the operations needed to apply one adiabatic step for the case of a $1 \times 2$ graphene lattice are summarized in Tab.~\ref{tab:step_trotter}, from top to bottom. Each row represents one application of the fermionic swaps. In each step, the qubits in red suffer an interaction corresponding to an horizontal edge in the lattice for qubits belonging to the same row. The qubits in green suffer an interaction corresponding to vertical edges. The full scheme covers all operations available in the $\H_H$ Hamiltonian. Observe that vertical interactions are skipped in some steps to accommodate missing lines in the honeycomb lattice. In addition, the middle row, which has two qubits more, needs two extra permutation steps. Those are taken after the first and last rows have completed half the permutation and their positions are inverted. In the case of the spin-orbit Hamiltonian $\H_{SO}$, an extra swap between $(2k, 2k + 1)$ is required at the beginning of the process in the odd rows. This adjusts the adjacent qubits to have different spins.

The circuit corresponding to an adiabatic step, for a system with $\lambda_R=0$, is depicted in Fig.~\ref{fig:gates_adiabatic}. Both the Coulomb and hopping terms are present in the circuit. The fermionic swaps modify the ordering of qubits in such a way that Hamiltonian evolution can be executed in a small subsets of qubits while affecting all of them. The particular value of $\theta$ inside the Hamiltonian operations will depend on the adiabatic evolution, being constant for the hopping term and increased in each step for Coulomb. 

\subsection{Measurement}

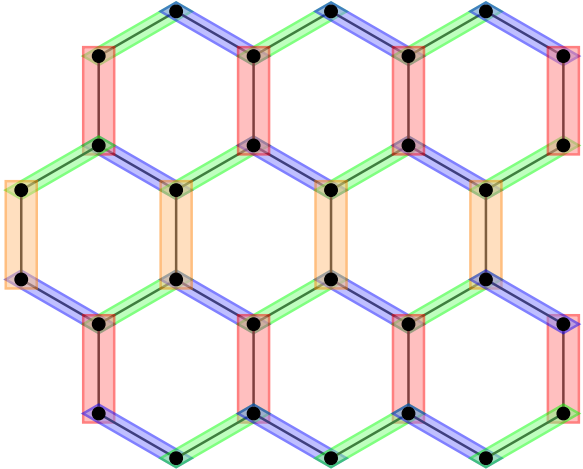
\begin{figure}[t]
    \resizebox{.9\linewidth}{!}{
    \begin{tikzpicture}
    \foreach \x in {0,1,2}{
    \draw[black, thick] (\x*1.732,0) -- (\x*1.732 + 0.866,1/2);
    \draw[green, thick, fill=green!50, opacity=0.5] (\x*1.732-0.1732,0) -- (\x*1.732,-0.1) -- (\x*1.732 + 0.866 + 0.1732,1/2) -- (\x*1.732 + 0.866,1/2 + 0.1) -- cycle;
    \draw[black, thick] (\x*1.732 + 0.866,1/2) -- (\x*1.732 + 1.732,0);
    \draw[blue, thick, fill=blue!50, opacity=0.5] (\x*1.732 + 0.866,1/2 + 0.1) -- (\x*1.732 + 0.866 - 0.1732,1/2) -- (\x*1.732 + 1.732,-0.1) -- (\x*1.732 + 1.732 + 0.1732,0) -- cycle;
    \draw[black, thick] (\x*1.732,-1) -- (\x*1.732 + 0.866,-3/2);
    \draw[blue, thick, fill=blue!50, opacity=0.5] (\x*1.732-.1732,-1) -- (\x*1.732,-1 + 0.1) -- (\x*1.732 + 0.866+.1732,-3/2) -- (\x*1.732 + 0.866,-3/2-0.1) -- cycle;
    \draw[black, thick] (\x*1.732 + 0.866,-3/2) -- (\x*1.732 + 1.732,-1);
    \draw[green, thick, fill=green!50, opacity=0.5] (\x*1.732 + 0.866 - .1732,-3/2) -- (\x*1.732 + 0.866,-3/2 - 0.1) -- (\x*1.732 + 1.732 + 0.1732,-1) -- (\x*1.732 + 1.732,-1 + 0.1) -- cycle;
    }
    \foreach \x in {0,1,2, 3}{
    \draw[black, thick] (\x*1.732,0) -- (\x*1.732,-1);
    \draw[red, thick, fill=red!50, opacity=0.5] (\x*1.732 - .1732,0 + 0.1) -- (\x*1.732 + .1732,0 + 0.1) -- (\x*1.732 + 0.1732,-1-0.1) -- (\x*1.732 - 0.1732,-1-0.1) -- cycle;
    };
    \draw[black, thick] (-0.866,-3/2) -- (0,-1);
    \draw[green, thick, fill=green!50, opacity=0.5] (-0.866-.1732,-3/2) -- (-0.866,-3/2-0.1) -- (0+.1732,-1)  -- (0,-1+.1) -- cycle;    
    \foreach \x in {0,1,2}{
    \draw[black, thick] (\x*1.732 - 0.866,-5/2) -- (\x*1.732,-3);
    \draw[blue, thick, fill=blue!50, opacity=0.5] (\x*1.732 - 0.866-.1732,-5/2) -- (\x*1.732 - 0.866,-5/2 + 0.1) -- (\x*1.732+.1732,-3) -- (\x*1.732,-3-0.1) -- cycle;
    \draw[black, thick] (\x*1.732,-3) -- (\x*1.732 + 0.866,-5/2);
    \draw[green, thick, fill=green!50, opacity=0.5] (\x*1.732-.1732,-3) -- (\x*1.732,-3 - 0.1) -- (\x*1.732 + 0.866+.1732,-5/2) -- (\x*1.732 + 0.866,-5/2+.1) -- cycle;}
    \foreach \x in {0,1,2, 3}{
    \draw[black, thick] (\x*1.732-.866,-3/2) -- (\x*1.732 - 0.866,-5/2);
    \draw[orange, thick, fill=orange!50, opacity=0.5] (\x*1.732-.866-.1732,-3/2+.1) -- (\x*1.732-.866+.1732,-3/2+.1) -- (\x*1.732 - 0.866+.1732,-5/2-.1) -- (\x*1.732 - 0.866-.1732,-5/2-.1) -- cycle;    
    }
    \foreach \x in {0,1,2, 3}{
    \draw[black, thick] (\x*1.732,-4) -- (\x*1.732,-3);
    \draw[red, thick, fill=red!50, opacity=0.5] (\x*1.732 - .1732,-4-.1) -- (\x*1.732 + .1732,-4-.1) -- (\x*1.732+.1732,-3 + .1) -- (\x*1.732-.1732,-3+.1) -- cycle;
    };
    \draw[black, thick] (5 * .866, -5/2) -- (6 * .866, -3);
    \draw[blue, thick, fill=blue!50, opacity=0.5] (5 * .866 - .173, -5/2) -- (5 * .866, -5/2 + .1) -- (6 * .866 + .173, -3) -- (6 * .866, -3 - .1) -- cycle;
    \foreach \x in {0,1,2}{
    \draw[black, thick] (\x*1.732,-4) -- (\x*1.732 + 0.866,-9/2);
    \draw[blue, thick, fill=blue!50, opacity=0.5] (\x*1.732 - .1732,-4) -- (\x*1.732,-4+0.1) -- (\x*1.732 + 0.866+.1732,-9/2) -- (\x*1.732 + 0.866,-9/2-0.1) -- cycle;
    \draw[black, thick] (\x*1.732 + 0.866,-9/2) -- (\x*1.732 + 1.732,-4);
    \draw[green, thick, fill=green!50, opacity=0.5] (\x*1.732 + 0.866 - .1732,-9/2) -- (\x*1.732 + 0.866,-9/2-.1) -- (\x*1.732 + 1.732+.1732,-4) -- (\x*1.732 + 1.732,-4 + 0.1) -- cycle;}
    \foreach \x in {0,2,4, 6}{
    \filldraw[black] (\x*.866,0) circle (2pt);
    };
    \foreach \x in {1,3,5}{
    \filldraw[black] (\x*.866,1/2) circle (2pt);
    };
    \foreach \l [count=\i, evaluate=\i as \x using int(2*\i - 2)] in {13,11,9, 7}{
    \filldraw[black] (\x*.866,-1) circle (2pt);
    };
    \foreach \l [count=\i, evaluate=\i as \x using int(2*\i - 3)] in {14, 12,10,8}{
    \filldraw[black] (\x*.866,-3/2) circle (2pt);
    };
    \foreach \l [count=\i, evaluate=\i as \x using int(2*\i - 3)] in {15, 17,19,21}{
    \filldraw[black] (\x*.866,-5/2) circle (2pt);
    };
    \foreach \l [count=\i, evaluate=\i as \x using int(2*\i - 2)] in {16,18,20, 22}{
    \filldraw[black] (\x*.866,-3) circle (2pt);
    };
    \foreach \l [count=\i, evaluate=\i as \x using int(2*\i - 2)] in {29,27,25, 23}{
    \filldraw[black] (\x*.866,-4) circle (2pt);
    };
    \foreach \l [count=\i, evaluate=\i as \x using int(2*\i - 1)] in {28, 26,24}{
    \filldraw[black] (\x*.866,-9/2) circle (2pt);
    };
\end{tikzpicture}}
\caption{Measurement scheme for the lattice. All terms from $H_{FH}$ coming from edges in the same color can be measured simultaneously. Horizontal terms can be measured directly since they do not commute. In the case of vertical terms, they are measured using swapping in a common substring in the full Jordan-Wigner mapping for all edges in the same row. Each of the four sets has two possible spins and regular and spin-orbit terms,
amounting to a total of 16 independent measurements.
Together with the Coulomb terms,
the complete measurement can be done in only 17 steps,
irrespective of the problem size. }
\label{fig:measurement}
\end{figure}

Once the adiabatic evolution is performed it is necessary to devise a measurement procedure to read out the energy of the system. A minimal set of measurements is obtained by grouping the different parts of the Hamiltonian into terms that commute with each other, so that they
can be measured simultaneously. For each term, a set of swapping gates is applied to bring the relevant qubits together, in case they are not. Then a change of basis is performed if the Hamiltonian term is non-diagonal.

All Coulomb terms commute with each other, are diagonal, and involve adjacent qubits. They are of the form $|11\rangle\langle 11|$, where both qubits represent the two spins in each site. The total Coulomb energy is the sum of probabilities of measuring 11, $P_{11}$, in the two qubits of each site times the coupling $U$.

In the case of hopping terms, non-adjacent qubits are involved and  swapping is needed. For horizontal terms, a set of even-odd and odd-even swaps for spins up and down suffices to bring both qubits together. The corresponding Hamiltonian term, $|XX\rangle\langle YY|$, is diagonalized to $|01\rangle\langle01|$ with the diagonalizing quantum circuit $\mathcal D = CNOT\; CH\; CNOT$, being $CH$ the control Hadamard two-qubit gate. The energy of each term is then $P_{10}-P_{01}$. Notice that the measurement strategy groups different 2-local Hamiltonian terms in such a way that all of them can be measured simultaneously, even though they do not commute given the Jordan-Wigner mapping.  

For vertical hopping terms the measurement must be done by taking many more qubits into account simultaneously. The grouping is done in even and odd rows of hexagons separately. Each pair of qubits in the vertical edge needs to be brought together to the right (or left) of the row, following the Jordan Wigner mapping and qubit chain. The number of swaps needed scales as $\mathcal{O}(N)$. The measurement diagonalization is then done as in the previous case, but it must be carried through all qubits in the corresponding row. 

Hopping spin-orbit Rashba terms are analogous to the regular kinetic terms, except for the presence of an additional complex phase. The swapping is adapted to involve qubits corresponding to different spins.

All sets of measurements add up to seventeen different groups, see Fig.~\ref{fig:measurement} for reference. For the Coulomb terms only one measurement is required. In the hopping case, only four different measurements are needed. Since there are two spins per electron, 8 terms are required. This same structure holds for the spin-orbit piece of the problem. The number of measurements corresponds to geometrical structures and does not depend on the size of the problem. 

\section{Simulation}\label{sec:simulation}

The problem of finding the ground state of the Hamiltonian $\H_{FH}$ from Eq.~\eqref{eq:h_fh} has been solved 
in this work with the algorithm previously exposed. However, there are different procedures to execute this algorithm depending on the available resources.

First of all, if the system is small enough it is possible to perform the exact diagonalization of the Hamiltonian to obtain its ground state. In this problem, the smallest possible system is one hexagon, corresponding to 12 qubits. In double precision, this state fits in a 68 kB memory. The Hamiltonian of this system requires 272 MB of memory to be stored in its dense form. This requirement is lowered by applying sparse matrices. In the case of two hexagons, memory requirements are 17 MB and 18 TB for statevector and Hamiltonian in dense form. This memory requirement is only achieved by supercomputers when running on large portions of the machine. In the computer MareNostrum4 hosted at BSC, the main memory of the full machine is only 324 TB~\cite{marenostrum}. For 3 hexagons, the memory requirements are 1.06 GB and $\sim 10^6$ TB, completely out of reach, even if using sparse matrices. For this reason, only exact diagonalizations up to 2 hexagons are available.  

The second simulation method is the adiabatic evolution by exact matrix exponentiation as described in Eq.~\eqref{eq:adiabatic}. This method is subject to the same memory limitations as the exact diagonalization.

Thus, both methods are used as benchmark tools for more sophisticated strategies that make use of the algorithm previously presented on efficient computers. The specific strategies developed for performing the simulations have been developed separately by the authors of this work.

\begin{figure*}[tbh]
    \centering
    \subfigure[~$C = \{0 \rightarrow 1 / 2\}; \lambda_R = 0$\label{fig:ev_coulomb}]{\includegraphics[width=.32\linewidth]{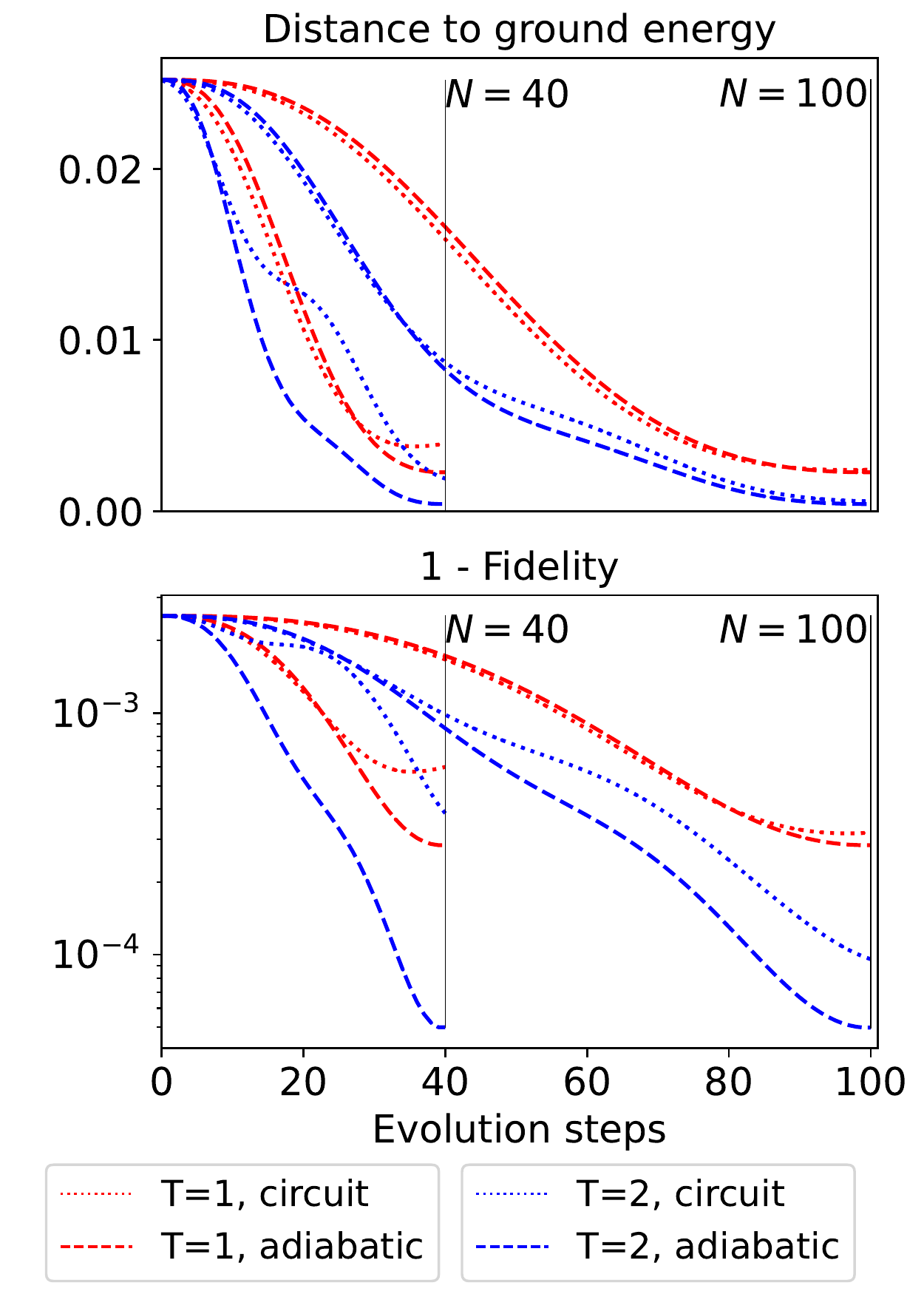}}
    \subfigure[~$C = \{0 \rightarrow 1 / 2\}; \lambda_R = 1$\label{fig:ev_rashba}]{\includegraphics[width=.32\linewidth]{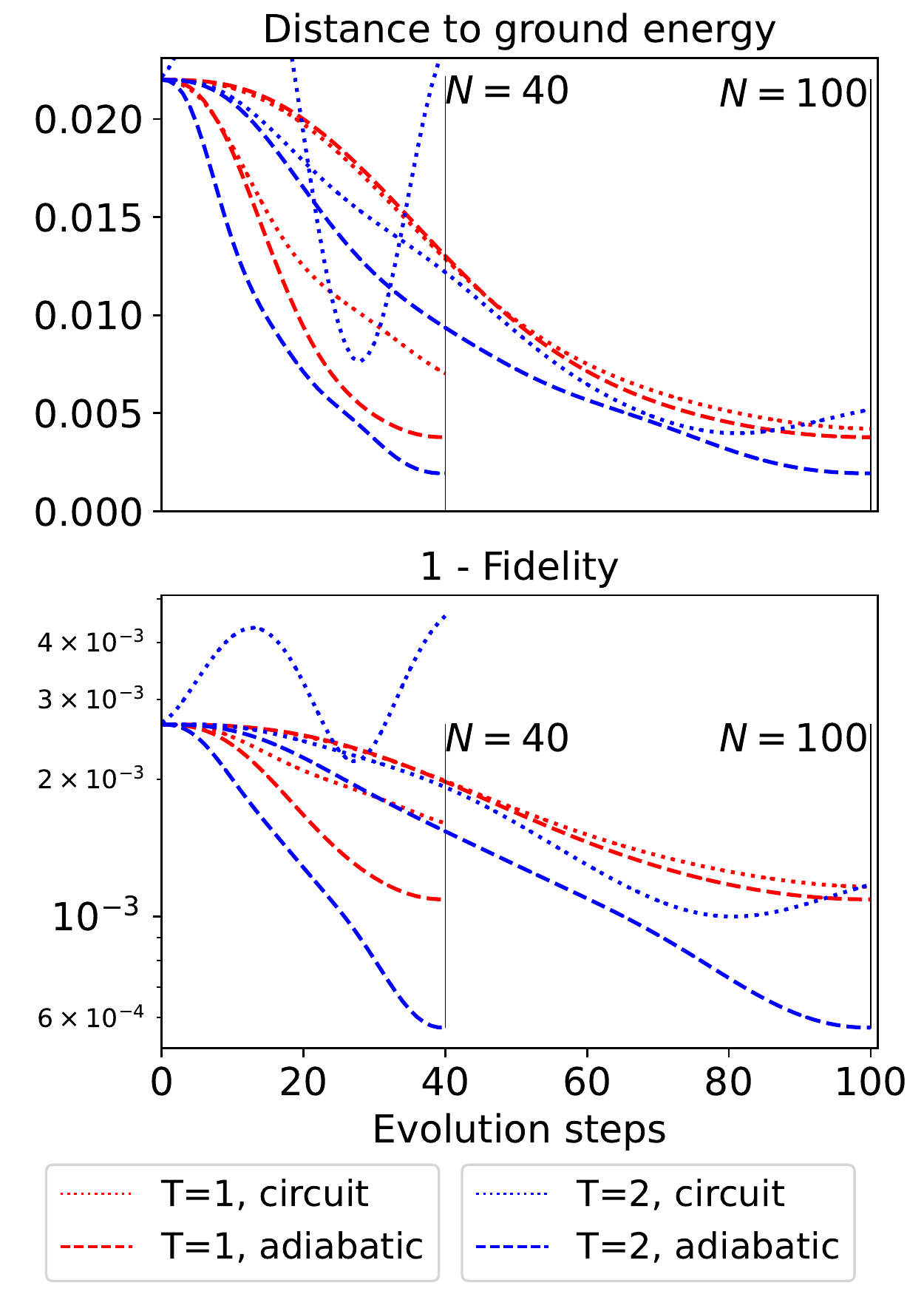}}
    \subfigure[~$C = \{0 \rightarrow 1 / 2\}; \lambda_R = \{0 \rightarrow 1\}$\label{fig:ev_rashba_SO}]{\includegraphics[width=.32\linewidth]{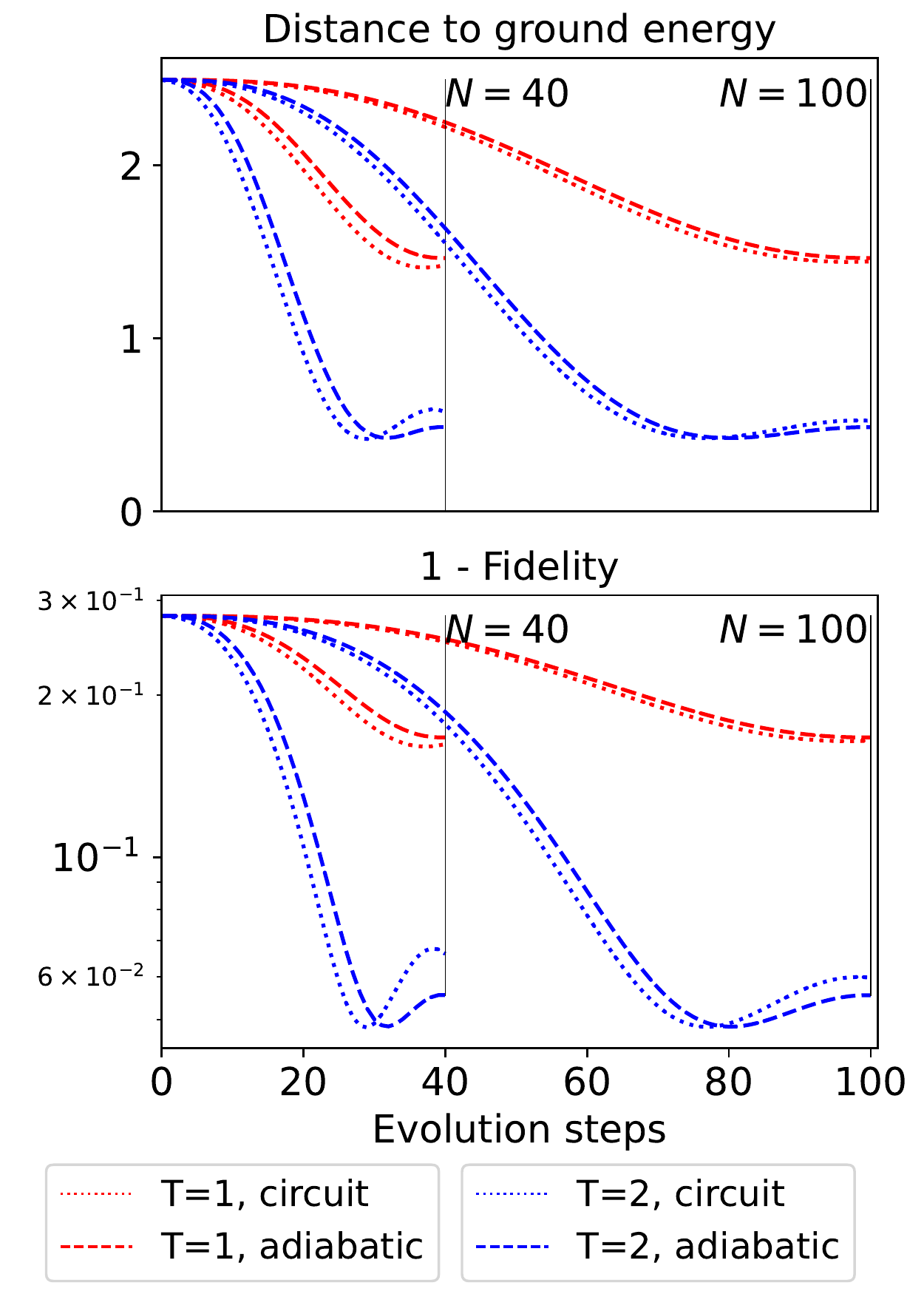}}
    \caption{Comparison between adiabatic evolutions, as described in Eq.~\eqref{eq:adiabatic}, and trotter evolutions, from Eq.~\eqref{eq:trotter}, for different values of $C$ and $\lambda_R$ detailed in each subcaption. All system are composed by one hexagon. For each example, comparison for two different periods $T$ and number of steps $N$ are displayed. The top figures represent the distance to the ground energy of the evolved state, while the bottom figures represent the fidelity of the evolved state with respect to the target one.}
    \label{fig:evolution_adiabatic_trotter}
\end{figure*}

The statevector simulation is used in this work to perform the circuit evolution of systems up to $2 \times 2$ hexagons, namely 32 qubits. Simulation is executed using the platform {\tt Qibo}~\cite{Efthymiou2021qibo}. This is a hardware-accelerated exact simulator for quantum circuits. In particular, through this work the JIT compilation supported by the {\tt Qibojit} backend and {\tt Numba}~\cite{Lam_numba_2015} is used to take advantage of several cores in parallel. The statevector in the hardest case needs 68 GB in double precision and 34 GB in single precision. The expected energy is computed in this case by storing two copies of the evolved statevector and modifying one according to the terms in the Hamiltonian to obtain the contribution of each one. This simulation is run on one node of the MareNostrum 4 machine, which has 48 parallel cores and 96 GB of memory.

An alternative method to perform simulation is the Matrix Product State~(MPS) approach. 
This method takes advantage of tensor networks tools, in which the statevector is defined as a tensor with as many indices as qubits, $\psi_{i_1,\cdots,i_{N_q}}$, with each index $i_j=0,1$ and yielding a total of $2^{N_q}$ values. In the MPS technique, this tensor is restructured into a product of $N_q$ tensors 
$\phi_{i_1}^{j1}\phi_{i_1,i_2}^{j_2}\phi_{i_2,i_3}^{j_3}\cdots\phi_{i_{N_q}}^{j_{N_q}}$.  Quantum gates operating on adjacent qubits are defined as tensors contracting adjacent $\phi$'s, thus translating the quantum circuit into a Tensor Network. The contractions in this tensor network can then be systematically approximated using singular value decomposition and cutting off the included singular values at a certain number, defined as the maximum bond dimension $\chi_M$. This quantity imposes a trade-off between memory required to store an approximation of the state, and the accuracy obtained. With this numerical approach, we are able to simulate with good accuracy lattices with 4 hexagons oriented vertically, encompassing a total of 36 qubits.

\section{results}\label{sec:results}

The first results here presented comprise the adiabatic evolution from the ground state of the tight binding lattice to the full FH model. As previously stated, this evolution can be simulated by applying evolution steps with the full Hamiltonian matrix for each time step, as in Eq.~\eqref{eq:adiabatic}, or by decomposing the matrix in simple terms corresponding to one- and two-qubit quantum gates, see Eq.~\eqref{eq:trotter}. The comparison between both methods for several examples is depicted in Fig.~\ref{fig:evolution_adiabatic_trotter}, for a one-hexagon lattice. In each case, two periods and two numbers of steps $N$ are considered. The top figures show the distance to the ground energy through the full process, while the bottom ones keep track of the fidelity with respect to the target state.

Fig.~\ref{fig:ev_coulomb} shows the evolution $C = 0 \rightarrow 1/2, \lambda_R = 0$ for periods $T=\{1, 2\}$ for $N=\{40,100\}$. By inspecting the adiabatic evolutions with different periods it is clear that this parameter affects directly the performance of the evolution. On the other hand, the number of steps in the adiabatic evolution does not play a relevant role. On the contrary, the Trotterization is highly sensitive to the number of steps. As this $N$ increases, the trotterized evolution simulates the adiabatic one more accurately. 

The same behavior is observed in Fig.~\ref{fig:ev_rashba} showing the evolution $C = 0 \rightarrow 1/2, \lambda_R = 1$. In this case, the Trotterization needs a higher number of steps to achieve accurate results. This is a direct consequence of having much more complex Trotter steps including hopping and also spin-orbit terms from the Hamiltonian. 

In Fig.~\ref{fig:ev_rashba_SO} the evolution $C = 0 \rightarrow 1/2, \lambda_R = 0 \rightarrow 1$ is depicted. In this case, the evolution becomes more complicated and requires higher periods to attain competitive results. This reflects the fact that the time-dependent adiabatic Hamiltonian has a lower gap between ground and first-excited state than the previous examples. In addition, the evolution shows an oscillatory behavior arising from the changes in the spin-orbit coupling while the hopping one remains equal. 

\begin{figure*}[tbh]
\centering
\subfigure[~${\rm Grid=1\times 1}$\label{fig:grid_coulomb_1x1}]{\includegraphics[height=.23\linewidth]{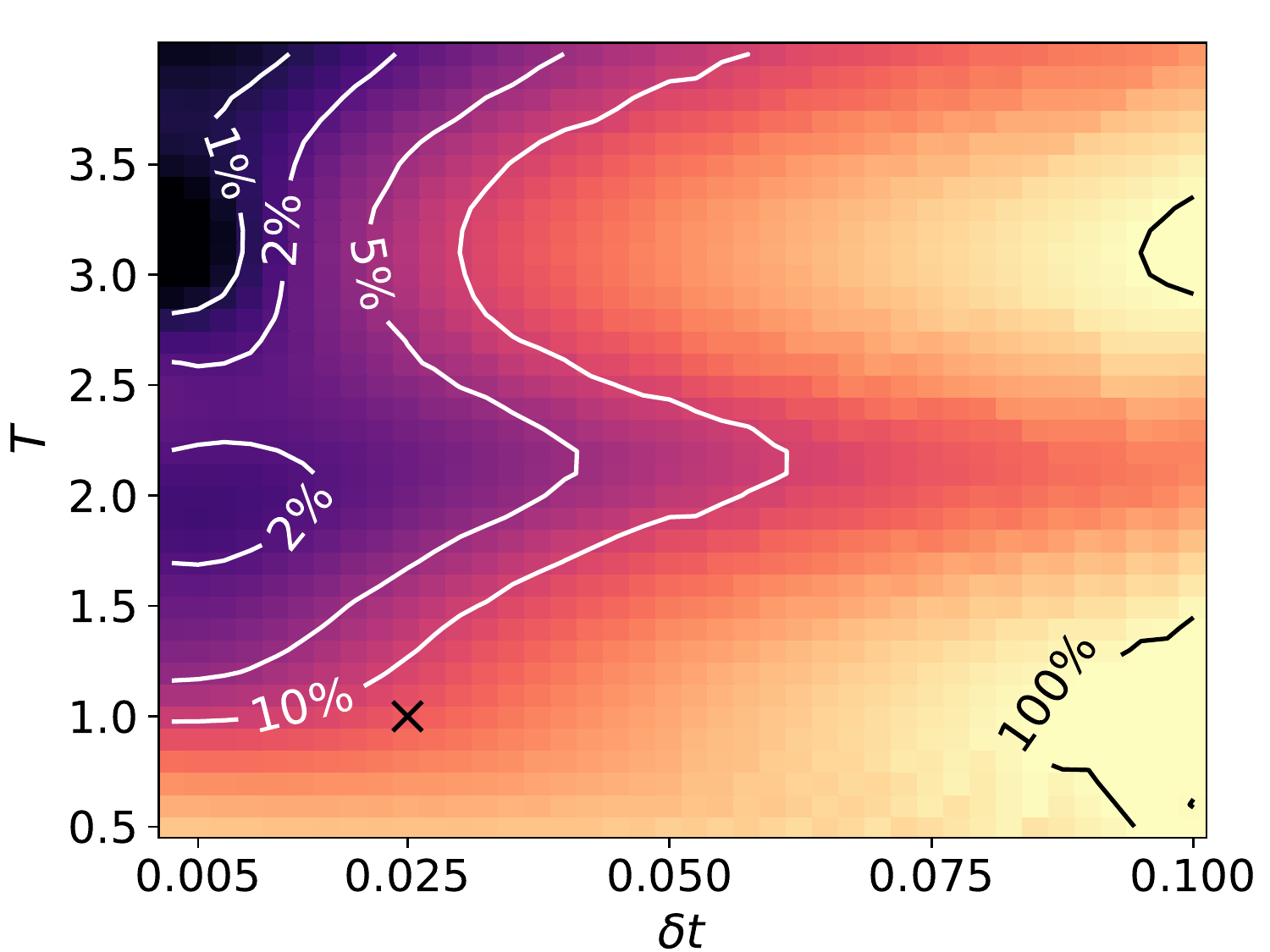}}
\subfigure[~${\rm Grid=2\times 1}$\label{fig:grid_coulomb_2x1}]{\includegraphics[height=.23\linewidth]{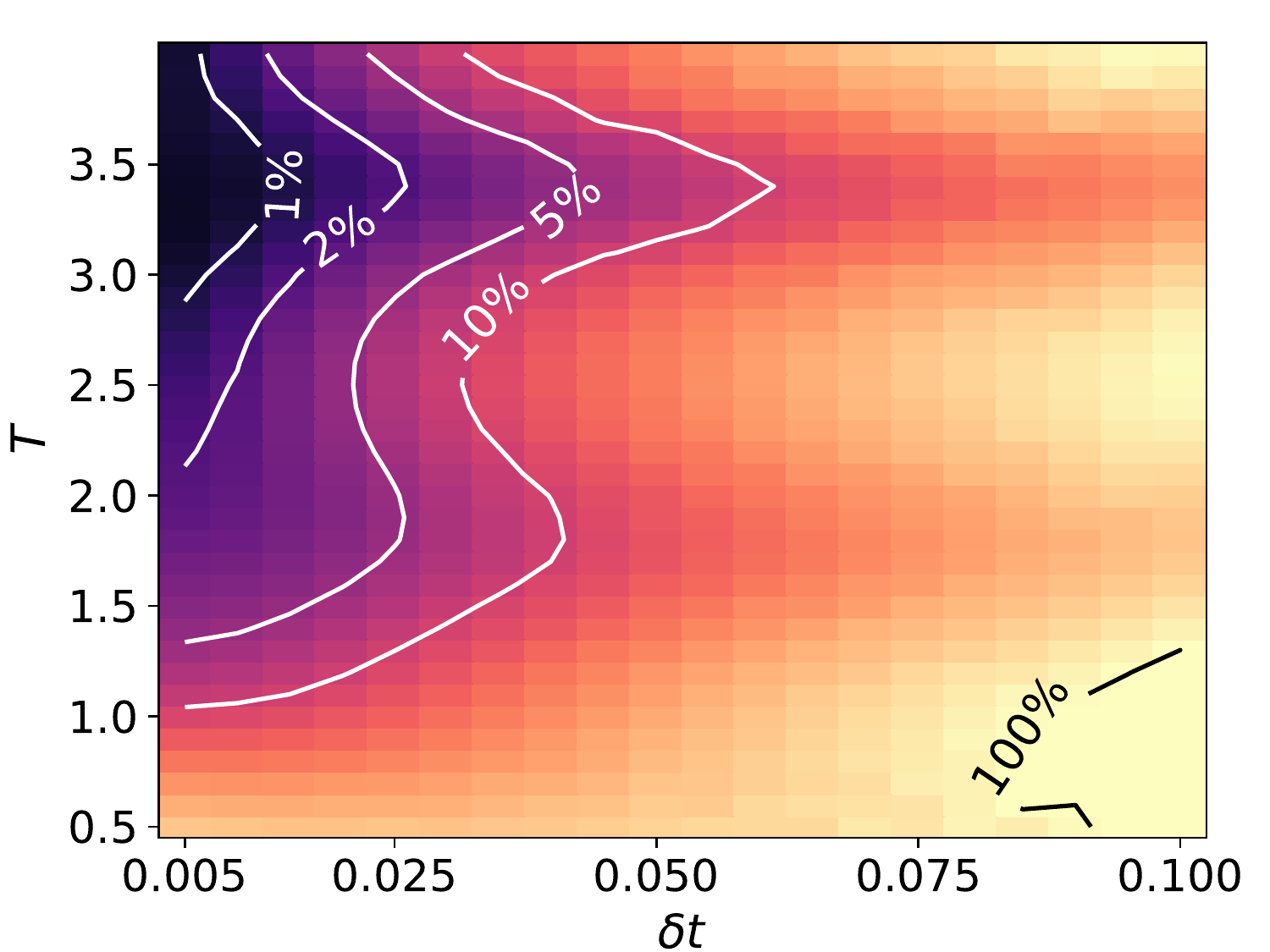}}
\subfigure[~${\rm Grid=1\times 2}$\label{fig:grid_coulomb_1x2}]{\includegraphics[height=.23\linewidth]{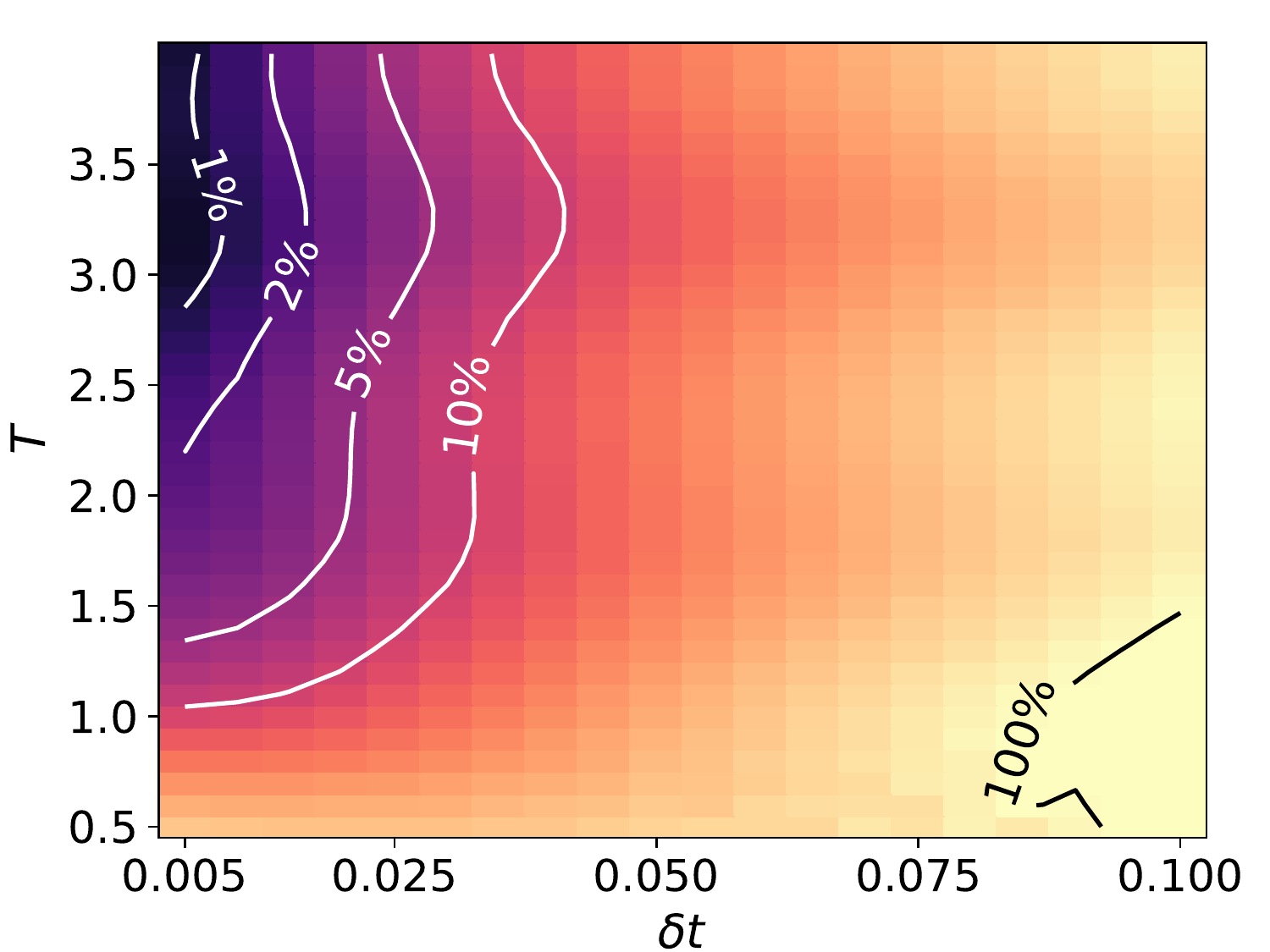}}
\subfigure[~${\rm Grid=1\times 1}$\label{fig:grid_rashba_1x1}]{\includegraphics[height=.23\linewidth]{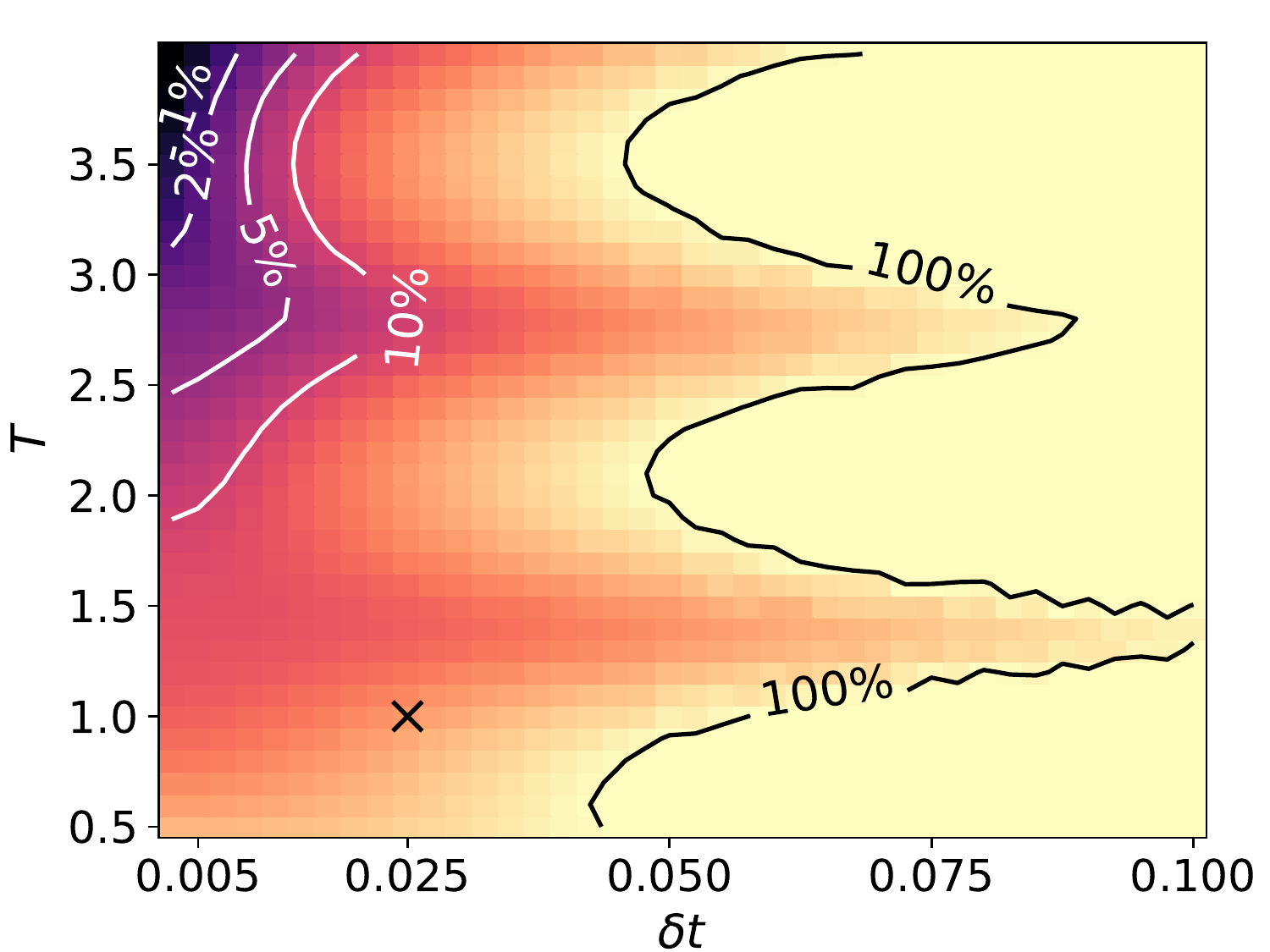}}
\subfigure[~${\rm Grid=2\times 1}$\label{fig:grid_rashba_2x1}]{\includegraphics[height=.23\linewidth]{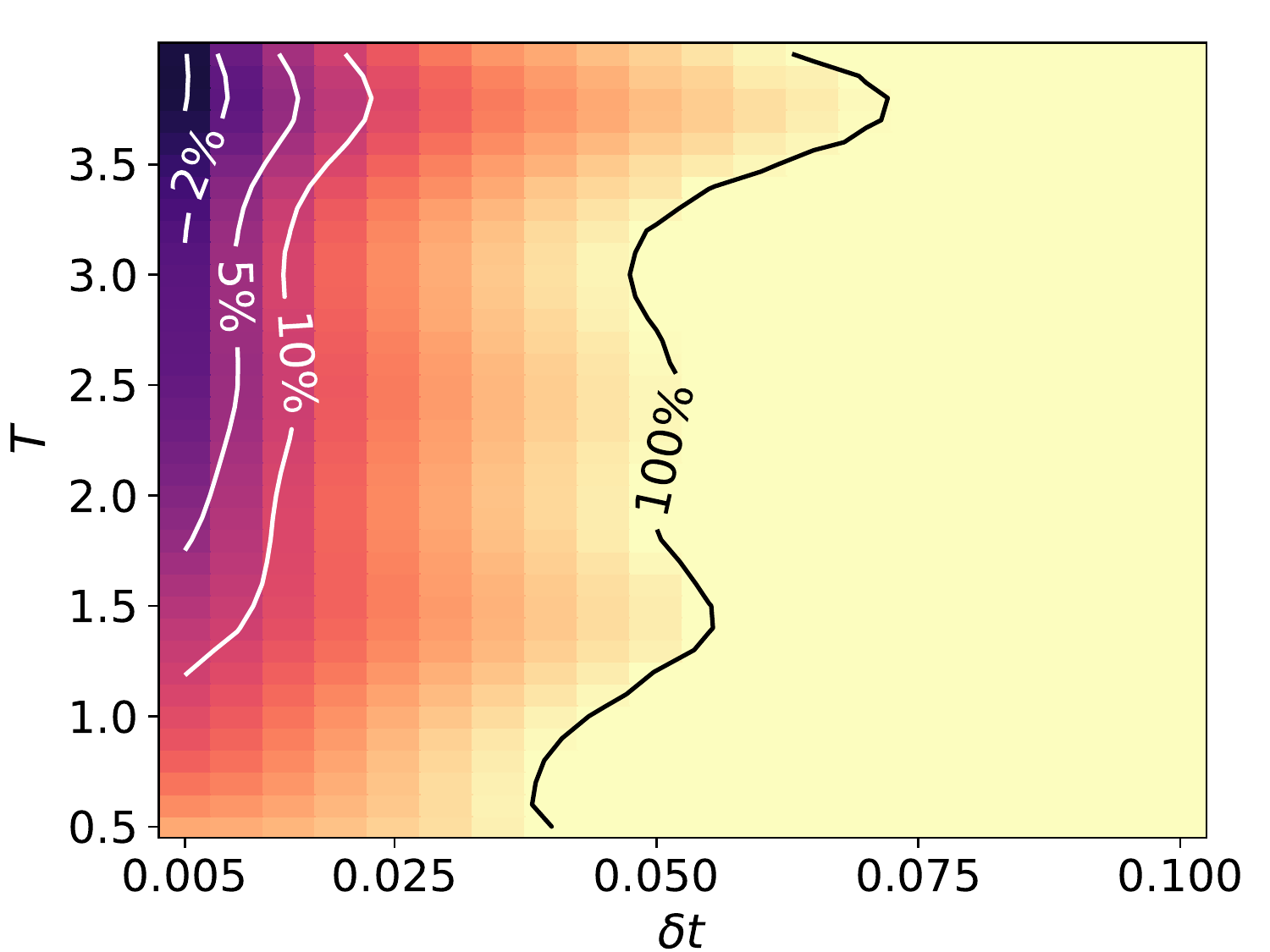}}  \hspace{.03\linewidth}
\subfigure{\includegraphics[height=.23\linewidth]{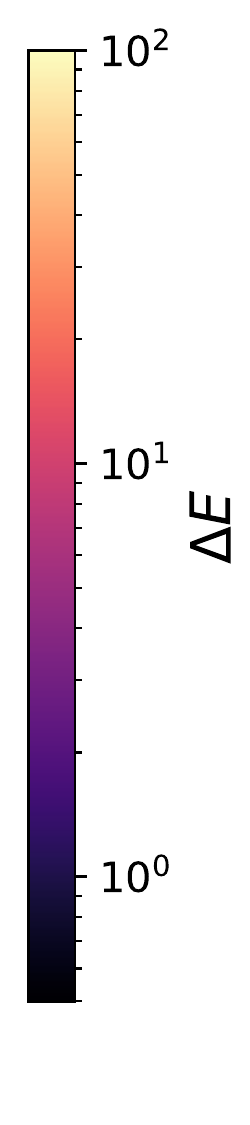}} \hspace{.22\linewidth}
\begin{textblock}{4}(10.1, -1.8)
\small \flushleft
Top row: \\
\hspace{3mm} $C = \{0 \rightarrow 1/2\}, \lambda_R = 0$ \\

\vspace{3mm} 

Bottom row: \\
\hspace{3mm} $C = \{0 \rightarrow 1/2\}, \lambda_R = 1$ \\
\end{textblock}
\caption{Error appearing in the circuit evolution for problems with $C = 1/2, \lambda_R = 0$ (top row) and $C = 1/2, \lambda_R = 1$ (bottom row) considering grids of different sizes. Evolution only contributes with Coulomb term. The marked $\times$ in (a) and (d) correspond to the $T=1, N=40,\, {\rm circuit}$ lines in Fig.~\ref{fig:evolution_adiabatic_trotter}. Errors above $100\%$ have been discarded from the plot. The error is measured taking as reference the difference between the ground energy and the starting one as defined in Eq.~\eqref{eq:delta_E}. This difference is typically $0.1$, while the target energy is $\sim 10$, depending on the size of the system. Thus, a $10\%$ error in the plot above corresponds to $1\%$ in the total energy. In general, different grids for the same problem parameters give rise to quantitatively similar error structures.}
\label{fig:grid}
\end{figure*}

It is important to compare Figs.~\ref{fig:ev_rashba} and~\ref{fig:ev_rashba_SO} since they are two different paths to arrive to the same goal. Using only a Coulomb evolution, triggers several advantages. First, the energy is always decreasing, thus it is guaranteed that the result will improve as more steps or larger times are considered. On the contrary, the full evolution presents unstable results. In addition, the Coulomb evolution in Fig.~\ref{fig:ev_rashba} starts closer to the target ground state both in terms of energy and fidelity, granting a better final result. For this reason, only Coulomb evolutions will be considered when presenting subsequent results. 

The next step is to scale the Trotterization for larger systems as it would be applied on an actual quantum computer. Fig.~\ref{fig:grid} shows the errors in the obtained energy for different examples depending on the period $T$ and time step $\delta t$. The error in the energy is here defined relative to the initial energy gap,
\begin{equation}\label{eq:delta_E}
    \Delta E = \left\vert \frac{\bra{\psi_{FH}} \H_{FH} \ket{\psi_{FH}} - \bra{\psi(t)} \H_{FH} \ket{\psi(t)}}{\bra{\psi_{FH}} \H_{FH} \ket{\psi_{FH}} -\bra{\psi_{TB}} \H_{FH} \ket{\psi_{TB}}} \right\vert.
\end{equation}

Figs.~\ref{fig:grid_coulomb_1x1}, \ref{fig:grid_coulomb_2x1}, \ref{fig:grid_coulomb_1x2} explore the $T, \delta t$ space for the example $C = 1, \lambda_R = 0$ in the lattices $(N_x, N_y) = \{(1,1),(2,1),(1,2)\}$. In this example a value of $\Delta E \approx 10\%$ is easily achieved for any $T>1$ and $\delta t < 0.05$, that is, using $N=20$ evolution steps. The $1\%$ threshold is obtained for $T>3$ and $\delta t < 0.025$, or $N = 120$. In all cases, a reasonable amount of steps of order $\sim 100$ suffices to get accurate predictions. Note also that in this example the $2\times 1$ grid returns better results than the physically equivalent $1 \times 2$, all error thresholds covering a larger area in the $2\times 1$ case. This can only arise from the algorithmic structure that involves different operator combinations. 

In Figs.~\ref{fig:grid_rashba_1x1}, \ref{fig:grid_rashba_2x1} the calculation is performed for the problem $C = 1/2, \lambda_R = 1$, for lattices $1 \times 1$ and $2 \times 1$. The starting point is the ground state of $C = 0, \lambda_R = 1$. In this case, the calculation is more costly than in cases with $\lambda_R = 0$ since the spin-orbit contribution must be added in each step. In addition, error values in this case are worse than with the previous case. To begin with, there is a large area where $\Delta E > 100\%$, rendering this set of parameters useless. In the $1\times 1$ case, the $\Delta E = 10\%$ threshold is located at $T>2.5, \delta t < 0.015 \rightarrow N = 167$. The $1\%$ threshold appears at $T>3.5, \delta t < 0.005; N = 700$. These results reflect also that simulating the spin-orbit is more sensitive to both adiabatic and Trotterization errors. To obtain comparable results, larger periods and shorter time steps are required, and therefore also many more steps $N$.

It is important to reflect the degree of approximation achieved with this method. The lowest errors obtained within these results gather at the order of $\Delta E \sim 1\%$. This includes the difference between the initial and final energies measured with the target Hamiltonian, which is usually at the order $E \sim 0.1$. 
Thus, the absolute error in the energy is of order $10^{-3}$, that as compared to the exact value returns a relative error of order $10^{-4}$.

Lattices with more than two hexagons are too computationally expensive to be solved with exact diagonalization. For larger systems we rely solely on the adiabatic evolution, which is subject to adiabatic and Trotter errors. The above analysis of these errors for one and two hexagons, together with convergence analysis when the number of steps and periods are increased, allow us to provide energy values for lattices with up to four hexagons with reasonable accuracy. Table~\ref{tab:large_lattices} shows these results for a fixed period $T=1$ and $N=40$ steps, corresponding to the points in Fig.~\ref{fig:grid}. Doubling these values shows converged results up to the third decimal digit, corresponding to relative errors as in Eq.~(\ref{eq:delta_E}) of $<20\%$, as expected from the full analysis in lattices with two hexagons. They are compared to the energies at time zero in the adiabatic evolution, that is the full FH energies given the ground state of the tight binding Hamiltonian.

For lattices with $N_q\leq32$ qubits, corresponding to SV occupying up to $34$ GB in single precision, we solve the adiabatic evolution exactly. For larger systems we  make use of structured tensor networks together with singular value decomposition.
This allows to trade computational cost with accuracy in the energies in a systematic way.
In particular, translating the circuit to an MPS and limiting the maximum bond dimension ($\chi_M$)
when contracting the tensors allows to reduce the memory needed while evaluating the circuit in reasonable times for lattices with up to 36 qubits. The accuracy of the state vector is given by the MPS fidelity, which is found very close to one or one. The case of four horizontal hexagons is an exception to this, yielding a relatively low fidelity $F=0.920$ for a maximum bond of 1000. This error is directly related to the non-locality of the Hamiltonian, which in this case contains a vertical hopping term operating on the first and last qubits, implying a separation of 32 qubits. This is then expected to worsen as the number of hexagons in the horizontal dimension increases. However, for $n\times 1$ lattices, a different Jordan-Wigner mapping may be used, in which the hexagons are oriented in an arm-chair way instead of a zig-zag one. This would benefit from the efficiency found in MPS simulations for $1\times n$ within the current Jordan-Wigner mapping.

\begin{figure}[t]
\centering
\includegraphics[width =.8\linewidth]{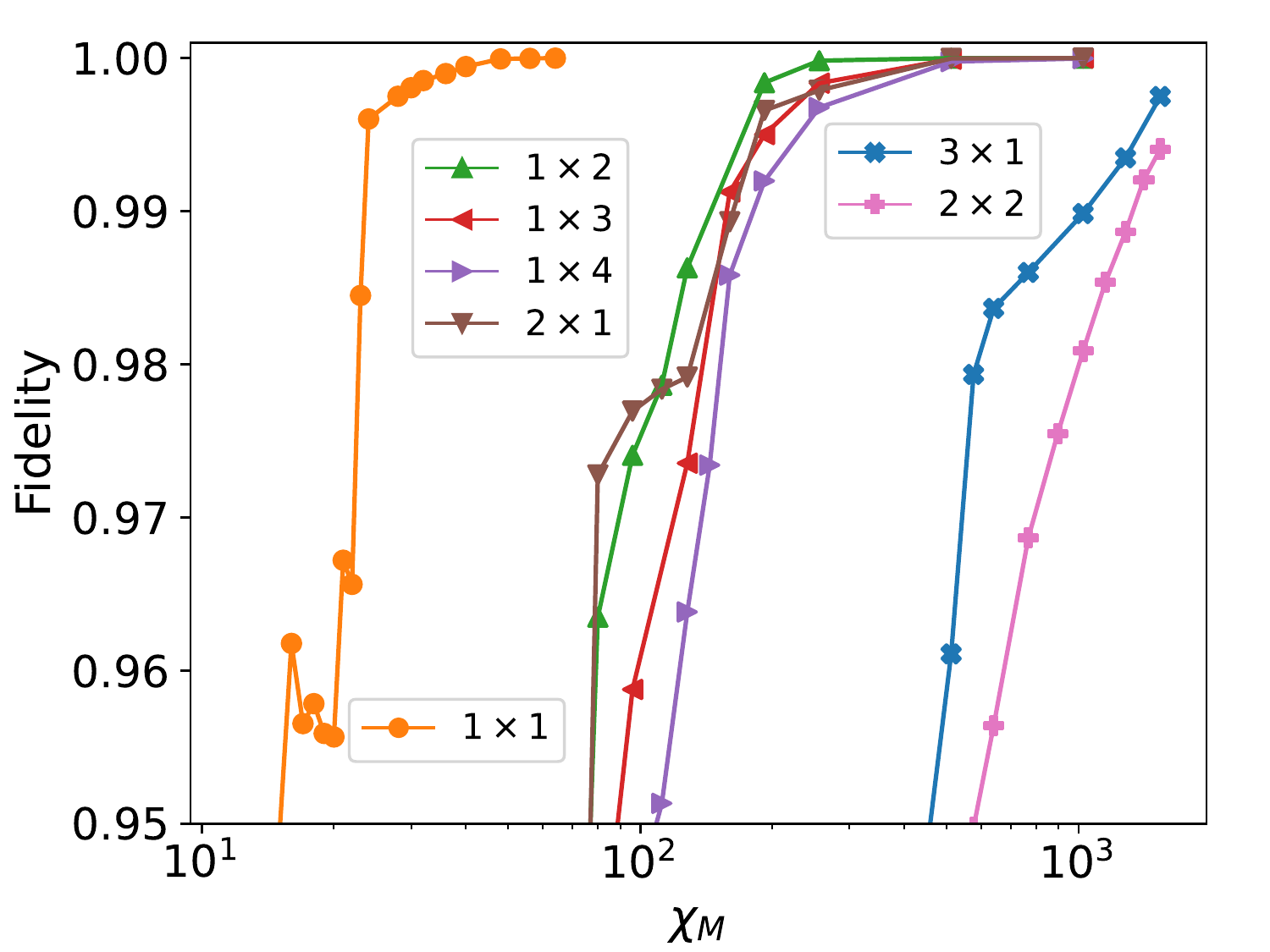}
\caption{MPS fidelity as a function of the maximum bond dimension $\chi_M$
for an adiabatic circuit with $N_s=40$, $T=1$, $U=0.5$, $\lambda_R=0$
for different lattices. Fidelity of the state vector is found to decay abruptly as $\chi_M$ is decreased passed a threshold. This threshold increases with the dimensions of the lattice, with the number of hexagons in the horizontal dimension playing the most relevant role.}
\label{fig:max_bond}
\end{figure}

The MPS method fidelity is analyzed for the lattices of Tab.~\ref{tab:large_lattices} and Fig.~\ref{fig:max_bond}. The lattice $4\times1$ requires large $\chi_M$ to obtain high enough fidelities. In fact, $E_T > E_0$, revealing that this $\chi_M$ is not enough to obtain better approximations than the initial state. For the other lattices, fidelity is found very close to one for bond dimensions much smaller
than the one yielding exact values, $\chi_e = 2^{N_q/2}$. These results indicate that the full FH ground states are not maximally entangled. In particular, Fig.~\ref{fig:max_bond} shows that $\chi_M$ can be decreased down to a certain value without any loss in the accuracy of the state vector. At a certain value, the fidelity decreases abruptly,
implying a finite entanglement entropy and that the statevector cannot be fully captured by further lowering the $\chi_M$.
This threshold value for $\chi_M$ roughly correlates with 
the locality of the corresponding Hamiltonian. In particular vertical hopping terms in the left of the lattice are less local and imply more entanglement as more horizontal hexagons are added to the lattice.

\begin{table}[t]
\begin{center}
\begin{tabular}{c|c|c|c|c|c c c}
 \multirow{2}{*}{\centering \textrm{lattice}} & \multirow{2}{*}{\centering $N_q$} & \multirow{2}{*}{\centering $E_{0}$} & \multicolumn{5}{c}{$E_{T}$}
 \\ \cline{4-8}
 & & & ED & SV & MPS & (F) & [$\chi_{M}$]
 \\ \hline
 $1\times1$ & 12 & -1.2083 & -1.2125 & -1.2118 & -1.2119 & (1.000) & [64]\\ \hline
 $2\times1$ & 20 & -1.2433 & -1.2474 & -1.2466 & -1.2467 & (1.000) & [512]\\ \hline
 $1\times3$ & 28 & -1.2641 & \ding{55} & -1.2674  & -1.2675 & (1.000) & [512]\\ \hline
 $3\times1$ & 28 & -1.2545 & \ding{55} & -1.2577 & -1.2571 & (0.997) & [1536]\\ \hline
 $2\times2$ & 32 & -1.2815 & \ding{55} & -1.2847 & -1.2842 & (0.994) & [1512]
 \\ \hline
 $1\times4$ & 36 & -1.2746 & \ding{55} &  \ding{55} & -1.2778 & (1.000) & [512]
 \\ \hline
 $4\times1$ & 36 & -1.2509 & \ding{55} & \ding{55} & -1.2445 & (0.920) & [1000] 
 \\ 
\end{tabular}
\end{center}
\caption{Ground energy per site of best estimates for different lattices and for $\lambda=0$, and $U=0$ and $U=0.5$, defined respectively as $E_0=\langle \psi(0)| H_{FH}|\psi(0)\rangle$ and
$E_T=\langle \psi(T)| H_{FH}|\psi(T)\rangle$. $N_q$ corresponds to the number of qubits of each lattice, ED for exact diagonalization, SV for state vector, and MPS for matrix product state. In the case of MPS, F stands for fidelity and $\chi_M$ for maximum bond dimension. SV has adiabatic and circuit Trotterization errors while MPS also suffers from truncation of $\chi_M$ errors.}
\label{tab:large_lattices}
\end{table}

\section{Conclusions}\label{sec:conclusions}

We have devised an algorithm to be run on a quantum computer that outputs the ground state and energy of artificial graphene given a set of qubits and a universal set of quantum gates.
The number of qubits corresponds to twice the number of sites, while the depth of the quantum circuit scales linearly with the size of the system.
This represents an exponential reduction in computational resources with respect to classical algorithms run on classical hardware.

The algorithm is based on a Trotterized adiabatic evolution, where an initial tight-binding ground state is prepared, and the Coulomb interaction is gradually tuned up.
Coulomb, hopping, and spin-orbit terms have been included, but the algorithm is flexible enough to include other interacting or spin-orbit terms that might be suitable for more general artificial graphene matter.

A full simulation of the algorithm has provided insight into what are the optimal periods and discretizations, two key inputs in the adiabatic evolution, and into the best adiabatic paths to the target state, were the algorithm to run in a quantum computer in large hexagonal lattices.
In particular, periods  $T\sim 1$, time steps $\delta t\sim 0.025$ generate results with $\sim 10\%$ errors for relatively large couplings,
and it is found optimal to prepare an initial state with all kinetic terms included.
The simulation yields accurate results for energies of lattices with up to four hexagons, or 36 qubits.

Different simulation methods have been used, each one capable of reaching a different maximum lattice size and subject to different sources of error.
Exact diagonalization is the most precise method but also the most limited, suitable for solving systems with at most two hexagons.
At the other extreme would be a quantum computer with dozens of error corrected qubits, which would be subject to adiabatic and discretization errors, but able to simulate AG with many hexagons.
An actual quantum computer would also suffer from decoherence and state fidelity quantum errors, which are not tackled in this work.
Between these two extremes we have simulated AG with adiabatic matrix multiplication, full statevectors, and MPS.

The adiabatic method errors depend only on the period and the size of the time step, while the state vector and MPS methods are also subject to errors due to Trotterization at each evolution step.
On the other hand, they allow to simulate systems with up to 32  and 36 qubits, respectively.
With MPS, we can trade off memory needs with fidelity of the state. Because the ground states found are not maximally entangled,
MPS has the potential to simulate larger systems without a relevant loss in fidelity of the state or accuracy in the energies.
In general, the exponential growth in classical resources implies simulations in current supercomputers are limited at the $\sim 30$s qubit barrier,
which corresponds to several hexagons in AG. 
In all cases explored, providing sufficient computational resources, errors in the final energy as compared to the initial energy below $1\%$ are possible. Initial energy is only slightly larger than the final one, so the relative error in the energy is below $10^{-4}$. 
This algorithm is then designed to provide very accurate descriptions of the ground state and of its energy. For the case of MPS, fidelities over $99\%$ are obtained.
Such high fidelities are possible because of the
large overlap between initial and target states,
allowing for a relatively short adiabatic path and correspondingly a very fine discretization.

With respect to increasing the range of applicability of this algorithm, further development in tensor network parallelization software such as {\tt RosNet} might be able to push this barrier to fairly larger systems.

\acknowledgements

We acknowledge to the team of {\tt Qibo} for their help when conducting simulations. This work has been financially supported by Secretaria d’Universitats
i Recerca del Departament d’Empresa i Coneixement
de la Generalitat de Catalunya, co-funded by the
European Union Regional Development Fund within
the ERDF Operational Program of Catalunya (project
QuantumCat, ref. 001-P-001644); and by the European Union Horizon 2020 research and innovation program under the Marie SklodowskaCurie grant agreement No. 754558 (PREBIST), and the AXA Chair in Quantum Information Science.

\bibliography{references.bib}

\end{document}